\documentclass[letterpaper,preprint,superscriptaddress,floatfix]{revtex4}
\usepackage[latin1]{inputenc}
\usepackage{multirow,amssymb,amsbsy,amsmath}
\usepackage{graphicx}
\usepackage{verbatim}
\makeatletter
\usepackage{pifont}
\makeatother
\usepackage{amsmath}
\usepackage{setspace}
\usepackage{bm}

\usepackage{siunitx}
\usepackage{xpatch}
\usepackage{caption}
\usepackage{ragged2e}

\newcounter{firstbib}

\begin{document}

\title{Heralded entanglement distribution between two absorptive quantum memories}

\author{Xiao Liu$\footnote{These authors contributed equally to this work\label{author}}$}
\author{Jun Hu\textsuperscript{\ref {author}}}
\author{Zong-Feng Li}
\author{Xue Li}
\author{Pei-Yun Li}
\author{Peng-Jun Liang}
\author{Zong-Quan Zhou$\footnote{email:zq\_zhou@ustc.edu.cn}$}
\author{Chuan-Feng Li$\footnote{email:cfli@ustc.edu.cn}$}
\author{Guang-Can Guo}
\affiliation{CAS Key Laboratory of Quantum Information, University of Science and Technology of China, Hefei, 230026, China}
\affiliation{CAS Center For Excellence in Quantum Information and Quantum Physics, University of Science and Technology of China, Hefei, 230026, China}
\date{\today}

\begin{abstract}
Owing to the inevitable loss in communication channels, the distance of entanglement distribution is limited to approximately 100 kilometres on the ground \cite{ED100KM}. Quantum repeaters can circumvent this problem by using quantum memory and entanglement swapping \cite{firstQP}. As the elementary link of a quantum repeater, the heralded distribution of two-party entanglement between two remote nodes has only been realized with built-in-type quantum memories \cite{QNcold1,QNcold2,QNcold4,QNion,QNatom,QNNV2,QNQD}. These schemes suffer from the trade-off between multiplexing capacity and deterministic properties and hence hinder the development of efficient quantum repeaters. Quantum repeaters based on absorptive quantum memories can overcome such limitations because they separate the quantum memories and the quantum light sources. Here we present an experimental demonstration of heralded entanglement between absorptive quantum memories. We build two nodes separated by 3.5 metres, each containing a polarization-entangled photon-pair source and a solid-state quantum memory with bandwidth up to 1 gigahertz. A joint Bell-state measurement in the middle station heralds the successful distribution of maximally entangled states between the two quantum memories with a fidelity of 80.4 $\pm$ 2.2 per cent ($\pm$1 standard deviation). The quantum nodes and channels demonstrated here can serve as an elementary link of a quantum repeater. Moreover, the wideband absorptive quantum memories used in the nodes are compatible with deterministic entanglement sources and can simultaneously support multiplexing, which paves the way for the construction of practical solid-state quantum repeaters and high-speed quantum networks.
\end{abstract}

\maketitle

Remote distribution of quantum entanglement and quantum states between any two locations is a fundamental task for large-scale quantum networks, which allow entanglement-based quantum key distribution \cite{qcommu}, distributed quantum computation, and enable large-scale tests of quantum physics \cite{QNNV2}. Photons are optimal carriers in quantum communication channels. However, direct transmission of photons through optical fibre is restricted to a few hundreds of kilometres because of the exponential channel loss \cite{reviewQP}. For long-distance entanglement distribution, one feasible solution is to transmit entangled photons over atmosphere channels through satellite-based links \cite{satellite}. In fibre-based telecommunications networks, one can divide the transmission channel into a number of elementary links and first create entanglement between two end nodes of each link. Then, through entanglement swapping between neighbouring links, entanglement can be gradually distributed to the target locations. This is known as the quantum repeater protocol \cite{firstQP,reviewQP}. To perform quantum communication tasks using quantum repeaters, the key building block is the generation of, in effect, two-excitation entanglement between remote quantum memories located at two nodes in a heralding way, that is, the realization of an elementary link for quantum repeaters. For this goal, entanglement generated from two-photon detections \cite{firstQP,reviewQP} is more directly applicable than number-state entanglement with single excitation, as it is difficult to perform qubit rotations in the number basis, and thus to make it functional the implementation of two chains of number-state entanglement between quantum memories in parallel is required \cite{QNcold1}. In addition, two-photon quantum interference greatly reduces the phase stability requirements for the channels, which is more robust and realistic for long-distance quantum communication \cite{reviewQP,twophoton,nonhiera}. 

Until now, the functional elementary link of a quantum repeater with heralded distribution of effective two-excitation entanglement has been achieved in cold atomic ensembles \cite{QNcold1,QNcold2,QNcold4} and in single quantum systems such as single trapped ions \cite{QNion}, single rubidium atoms \cite{QNatom}, nitrogen-vacancy centres \cite{QNNV2} and quantum dots \cite{QNQD}. Nevertheless, all of these demonstrations are based on built-in-type quantum memories (or emissive quantum memories) and atom-photon entanglement. For cold atomic ensembles, the atom-photon entanglement is generated between the collective excitation of an atomic ensemble and a Raman scattered photon, which is intrinsically probabilistic. For single quantum systems, although the entanglement can be deterministically generated between the atomic spin and an emitted photon, it is difficult to realize multiplexing. Schemes using absorptive quantum memories can overcome these issues (see Fig. 1). This type of quantum memory is flexible with the choice of entanglement sources, including deterministic entangled photon pairs, while retaining the capability of multiplexed operations \cite{multiQP}, and therefore should be more efficient for quantum repeater applications.

Here we experimentally demonstrate the heralded distribution of two-excitation entanglement between absorptive quantum memories based on rare-earth-ion-doped crystals. The solid-state quantum memories have shown attractive characteristics, such as long lifetime \cite{spinwave15,sixhours}, large bandwidth \cite{band16g}, high fidelity \cite{highF,highF2} and high multimode capability \cite{nonhiera,mode2,mode1,mode3}, which make it one of the most promising candidates for realizing quantum repeaters. In this proof-of-principle demonstration, we use an entangled-photon-pair source generated from spontaneous parametric down-conversion (SPDC). The experimental setup is illustrated in Fig. 2. Our experiment consists of two similar quantum repeater nodes. In each node, photon pairs are generated through the type-II SPDC process by pumping a nonlinear waveguide. To generate polarization entanglement states, the photon pairs are then sent to an interferometer after being spectrally filtered to a linewidth of approximately 1 GHz. Neglecting the multi-pair emissions, the state of photon pairs can be prepared to a nearly maximally entangled state $\vert\Phi^{+}\rangle=\left(1/\sqrt{2}\right)\left(\vert H\rangle\vert H\rangle+\vert V\rangle\vert V\rangle\right)$ through post-selection of coincidence counts between two outputs of the interferometer, where $\vert H\rangle$ ($\vert V\rangle$) denotes the horizontal (vertical) polarization states of the photons.

In our experiment, the second-order cross-correlation function between photon pairs at zero time-difference, $g_{12}^{(2)}(\Delta\tau)$, is set to 50(4) (all errors herein are one standard deviation) in a detection window of $\Delta\tau=2$ ns for a balance between brightness and non-classical correlation of the photon-pairs (see Methods and Extended Data Fig. 1). The counting rate is approximately 6,000 s$^{-1}$ GHz$^{-1}$. In order to fully characterize the two-photon polarization entanglement, we perform quantum state tomography on two sources, denoted as $S_{A}$ in node A and $S_{B}$ in node B, with the result shown in Fig. 3. We obtain a fidelity of 92.6(3)\% and 93.3(3)\% for $S_{A}$ and $S_{B}$, respectively.

Two quantum memories are placed in separated cryostats located at corresponding nodes near the entanglement sources. We use the atomic frequency comb (AFC) protocol \cite{AFC} for the storage of entangled photons. The absorption of a photon leads to collective excitations of an ensemble of ions, the inhomogeneously broadened absorption line of which has been tailored into a series of comb-shaped absorption peaks by spectral hole burning. After a programmed time that is associated with the period of absorption peaks, a rephasing occurs and the photon will be released. Once a photon from an entangled photon pair is stored in the quantum memory, its atomic excitations become entangled with the other photon, yielding the state $\left(1/\sqrt{2}\right)\left(\vert H\rangle\vert H_{M}\rangle+\vert V\rangle\vert V_{M}\rangle\right)$, where $\vert H_{M}\rangle$ ($\vert V_{M}\rangle$) represent the states of the quantum memory with H- (V-)polarized photon-induced excitations. Each memory hardware is composed of two pieces of Nd$^{3+}$:YVO$_{4}$ crystals sandwiching a half-wave plate (HWP) orientated at $45^{\circ}$ (see Methods). This compact device has been previously used for the storage of photonic polarization qubits with a fidelity above 99\% \cite{highF}.

We then quantify the performance of quantum memories by independently storing entangled photons from $S_{A}$ and $S_{B}$. Fig. 4a shows the photon-counting histogram of $S_{A}$ before and after storage of photons in the quantum memory in node A. The intrinsic storage efficiency is 14.3(1)\% in a 3-ns detection window for a storage time of 55.6 ns, which can be prolonged to over 1,250 ns with the sacrifice of efficiency (see Methods). A tomographic result of $S_{A}$ with photons retrieved from the quantum memory after 55.6 ns is shown in Fig. 4b, yielding a fidelity of 95.7(8)\% with respect to $\vert\Phi^{+}\rangle$. The memory fidelity between input and output states is 96.6(9)\%. For node B, the storage efficiency is 12.5(1)\% and the memory fidelity is 97.0(4)\%. The results indicate that entanglement preserves well in the quantum memories.

To demonstrate heralded entanglement distribution between two nodes, one entangled photon from each node (photons 2 and 3) are sent to an intermediate station for joint Bell-state measurement (BSM) through 5-m fibres while the other two photons (photons 1 and 4) are stored in two quantum memories separated by 3.5 m, as depicted in Fig. 2. Considering effective terms during the fourfold coincidence measurement of photons 1, 2, 3 and 4 (see Methods), the quantum state of the whole system before photons interacting on the BSM can be written as
\begin{equation}\label{whole}
\begin{aligned}
\vert\psi\rangle_{AB}=&\dfrac{1}{2}\left(\vert H\rangle_{2}\vert H_{M}\rangle_{1}+\vert V\rangle_{2}\vert V_{M}\rangle_{1}\right)\otimes\left(\vert H\rangle_{3}\vert H_{M}\rangle_{4}+\vert V\rangle_{3}\vert V_{M}\rangle_{4}\right)\\
=&\dfrac{1}{4}(\vert\Phi^{+}\rangle_{23}\vert\Phi^{+}_{M}\rangle_{14}+\vert\Phi^{-}\rangle_{23}\vert\Phi^{-}_{M}\rangle_{14}+\vert\Psi^{+}\rangle_{23}\vert\Psi^{+}_{M}\rangle_{14}+\vert\Psi^{-}\rangle_{23}\vert\Psi^{-}_{M}\rangle_{14}).
\end{aligned}
\end{equation}
where $\otimes$ denotes the Kronecker product. Here, $\vert\Phi^{\pm}\rangle_{23}=\left(1/\sqrt{2}\right)\left(\vert H\rangle_{2}\vert H\rangle_{3} \pm \vert V\rangle_{2}\vert V\rangle_{3}\right)$ and $\vert\Psi^{\pm}\rangle_{23}=\left(1/\sqrt{2}\right)\left(\vert H\rangle_{2}\vert V\rangle_{3}\pm\vert V\rangle_{2}\vert H\rangle_{3}\right)$ are four maximally-entangled Bell states for the photon 2 and 3 before BSM. The atomic states of quantum memories originated from the storage of photon 1 and 4 are expressed in similar forms as $\vert\Phi^{\pm}_{M}\rangle_{14}=\left(1/\sqrt{2}\right)\left(\vert H_{M}\rangle_{1}\vert H_{M}\rangle_{4} \pm \vert V_{M}\rangle_{1}\vert V_{M}\rangle_{4}\right)$ and $\vert\Psi^{\pm}_{M}\rangle_{14}=\left(1/\sqrt{2}\right)\left(\vert H_{M}\rangle_{1}\vert V_{M}\rangle_{4}\pm\vert V_{M}\rangle_{1}\vert H_{M}\rangle_{4}\right)$. The BSM is performed using linear optical elements such as polarizing beamsplitters (PBSs) and HWPs oriented at \ang{22.5} and subsequent single-photon detectors. A valid twofold coincidence counting of four outputs of BSM in a 2-ns detection window projects the states of photons 2 and 3 at the intermediate station onto $\vert\Phi^{+}\rangle_{23}$, which heralds the distribution of entangled state $\vert\Phi^{+}_{M}\rangle_{14}$ between two remote quantum memories.

Subsequently, we verify the entanglement established between two quantum memories by measuring the entangled photon pair (photons 1 and 4) converted back from the atomic excitations. To demonstrate the existence of entanglement between them, we determine the directly measurable observables called entanglement witnesses \cite{qw2}, which have positive expectation values on all separable states. Thus, a state is proved to be entangled if we observe negative expectation values when detected by witnesses. In our experiment, we use a witness given by
\begin{equation}\label{witness}
\begin{aligned}
\mathcal{W}=&\dfrac{1}{2}(\vert HV\rangle\langle HV\vert+\vert VH\rangle\langle VH\vert+\vert DA\rangle\langle DA\vert \\
&+\vert AD\rangle\langle AD\vert-\vert RL\rangle\langle RL\vert-\vert LR\rangle\langle LR\vert).
\end{aligned}
\end{equation}
Here $\vert D\rangle=\left(\vert H\rangle+\vert V\rangle\right)/\sqrt{2}$ and $\vert A\rangle=\left(\vert H\rangle-\vert V\rangle\right)/\sqrt{2}$ denote diagonal and antidiagonal polarization states, and $\vert R\rangle=\left(\vert H\rangle-i\vert V\rangle\right)/\sqrt{2}$ and $\vert L\rangle=\left(\vert H\rangle+i\vert V\rangle\right)/\sqrt{2}$ represent the right and left circular polarization states. Three sets of polarization basis are used for local measurements on individual photons following the operators defined above, with the results shown in Fig. 5. Let us assume that the density matrix of the measured entanglement state is $\rho_{m}$; we then obtain $\langle \mathcal{W}\rangle={\rm Tr}\left(\mathcal{W}\rho_{m}\right)=-0.304\pm0.022$ ($\pm$1 standard deviation), which is 14 standard deviations below zero and thus unambiguously proves the presence of bipartite entanglement between the two quantum memories.

Furthermore, we quantify the quality of our demonstration by checking the fidelity of the entanglement state achieved between the two quantum memories with respect to the ideal state, as defined by 
\begin{equation}\label{fidelity}
\mathcal{F}={\rm Tr}\left(\rho_{m}\vert\Phi^{+}_{M}\rangle_{14} \langle \Phi^{+}_{M}\vert \right),
\end{equation}
which can also be determined by the same measurement settings as used in the quantum witness detection. To do this, we can write the density matrix of the desired ideal state of retrieved photons as
\begin{equation}\label{densitymatrix}
\vert\Phi^{+}\rangle_{14} \langle \Phi^{+}\vert=\dfrac{1}{4} \left( I+\hat{\sigma}_{x}\hat{\sigma}_{x}-\hat{\sigma}_{y}\hat{\sigma}_{y}+\hat{\sigma}_{z}\hat{\sigma}_{z} \right),
\end{equation}
where $\hat{\sigma}_{x}=\vert D\rangle\langle D\vert-\vert A\rangle\langle A\vert$, $\hat{\sigma}_{y}=\vert R\rangle\langle R\vert-\vert L\rangle\langle L\vert$ and $\hat{\sigma}_{z}=\vert H\rangle\langle H\vert-\vert V\rangle\langle V\vert$ are three Pauli matrices. This means that we can extract the fidelity by performing three local measurements $\hat{\sigma}_{x}\hat{\sigma}_{x}$, $\hat{\sigma}_{y}\hat{\sigma}_{y}$ and $\hat{\sigma}_{z}\hat{\sigma}_{z}$ on the two retrieved photons. According to the measurement results in Fig. 5 and Eq. (\ref{fidelity}), the measured fidelity for the heralded entanglement is $\mathcal{F}=(80.4\pm2.2)\%$. The fidelity of retrieved two-photon entanglement is robust to the photon loss and the background noise, and the primary limit is from multi-pair emissions of the sources (see Methods). By reducing multi-photon emissions, we can see the increasing fidelity of the heralded two-photon entanglement before storage and observe the experimental violations of Bell's inequality. The highest $S$ parameter we obtained was $S=2.43\pm0.13$, a violation of Bell's inequality by more than 3 standard deviations (see Methods and Extended Data Fig. 2).

The entanglement distribution rate (EDR) achieved in our demonstration--that is, the final fourfold coincidence rate after detection by a superconducting nanowire single-photon detector (SNSPD)--is approximately 1.1 h$^{-1}$ with an average heralding rate of 100 Hz (see Methods). The EDR is primarily limited by the probabilistic generation of entangled photon pairs through the SPDC process. The use of a deterministic entanglement source--for example, from cascade emission of quantum dots \cite{QDE4}--together with high-density multiplexing, would greatly improve EDR in our setup with absorptive quantum memories (see Methods), leading to direct applications in some quantum repeater protocols with non-hierarchical architecture \cite{nonhiera}, in which the quantum memories only require preprogrammed storage times. In our demonstration, 55.6 ns/12.5 ns $\approx 4$ distinguishable temporal modes--spaced by 12.5 ns, as determined by the repetition rate of the pump laser--are actually stored in the quantum memories simultaneously thanks to the intrinsic temporal multimodality of the AFC protocol \cite{AFC}; this number can be further increased to approximately 56 by fully using the available time-bandwidth product ($55.6\ {\rm ns}\times 1\ {\rm GHz}$) (see Methods and Extended Data Fig. 5). The EDR can increase linearly with the number of modes. More intriguingly, Nd$^{3+}$:YVO$_{4}$ crystals used in this work have shown the capability of storage of 100 temporal modes of deterministic single photons generated from a quantum dot \cite{mode2} and multiplexed storage of 51 spatial modes \cite{mode1}, as well as easy integrations \cite{highF2}. Multiplexed storage using frequency \cite{nonhiera} and multiple degrees of freedom \cite{mode3} have also been demonstrated in similar materials. On-demand readout can be realized by adapting the two-level AFC to the Stark-modulated AFC scheme \cite{eleccontrol2} or using odd isotopes of Nd$^{3+}$ to achieve long-lived spin-wave storage \cite{AFC,NdEPR}, which would make this device more versatile for the practical applications. In addition, cavity enhancement can be employed to enhance the storage efficiency to close to unity \cite{cavity1}. To extend the elementary links to long distances, frequency conversion to telecom wavelength is needed. Thanks to the great wavelength flexibility of absorptive quantum memories, alternative approaches can be adopted, such as the use of photon-pair sources with one photon compatible with the quantum memory and the other photon at telecom wavelengths, or the use of erbium-doped crystal \cite{Er3} with a working wavelength near 1,550 nm.

To conclude, we have successfully distributed the heralded high-fidelity entanglement between two remote nodes through entanglement swapping in the intermediate station. This demonstration is, to our knowledge, the first physical realization of an elementary link for a quantum repeater based on absorptive quantum memories and entangled-photon-pair sources. Our configuration is compatible with deterministic entanglement sources and multiplexed operations, which are crucial for enhancing EDR, and thereby bringing us one step closer to efficient and high-speed quantum repeater architecture. Furthermore, with the above-mentioned future improvements, large-scale quantum networks can be constructed with the quantum repeater relying on absorptive solid-state quantum memories.

~\\

\noindent{\bf  Acknowledgements}
This work is supported by the National Key R\&D Program of China (No. 2017YFA0304100), the National Natural Science Foundation of China (Nos. 11774331, 11774335, 11504362, 11821404 and 11654002), Anhui Initiative in Quantum Information Technologies (No. AHY020100), the Key Research Program of Frontier Sciences, CAS (No. QYZDY-SSW-SLH003), the Science Foundation of the CAS (No. ZDRW-XH-2019-1), and the Fundamental Research Funds for the Central Universities (No. WK2470000026, No. WK2470000029 and No. WK2030000022). Z.-Q.Z. acknowledges the support from the Youth Innovation Promotion Association CAS.

\noindent{\bf Author Contributions}
Z.-Q.Z. and C.-F.L. designed the experiment. X.Liu and J.H. carried out the experiment with assistance from Z.-F.L. and X.Li. P.-Y.L. and P.-J.L. helped collect the data. X.Liu, J.H. and Z.-Q.Z. analysed the data and wrote the paper with input from all other authors. The project was supervised by Z.-Q.Z., C.-F.L. and G.-C.G. All authors discussed the experimental procedures and results.

\noindent{\bf Competing interests}
The authors declare no competing interests.

\clearpage

\begin{figure}[tb]
\centering
\includegraphics[width= 0.7 \columnwidth]{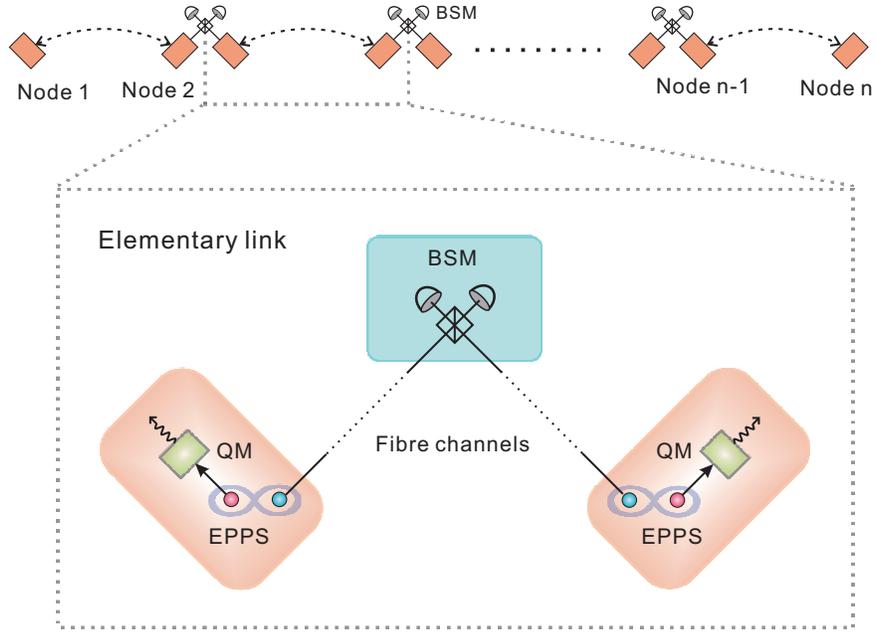}
\caption*{\justifying \textbf{Fig. 1: Schematic diagram for a quantum repeater and an elementary link with absorptive quantum memories.} In a quantum repeater, a long distance can be divided into several nodes linked by fibre channels. For an elementary link, two nodes are located at the ends, each containing a quantum memory (QM) and an external entangled-photon-pair source (EPPS). For each photon pair, one of them is stored in a quantum memory while the other is transmitted to the middle station over a fibre channel to be measured by Bell-state measurement (BSM). A click of the BSM heralds the successful distribution of entanglement between the two quantum memories. The photons stored in the quantum memories are then released for the next process of the quantum repeater, that is, the entanglement swapping process between different links to extend the entanglement to a longer distance.}\label{SYT}
\end{figure}

\begin{figure}[tb]
\centering
\includegraphics[width=1 \columnwidth]{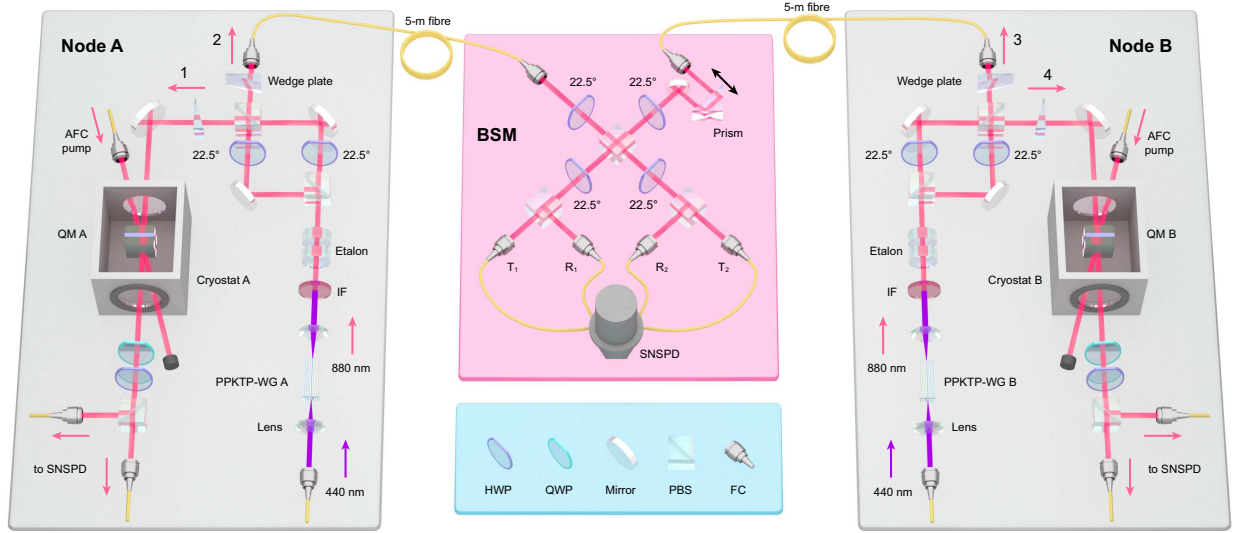}
\caption*{\justifying \textbf{Fig. 2: Experimental setup.} The whole demonstration involves two similar quantum repeater nodes (node A and node B) and a middle station for Bell-state measurement (BSM). In each node, a periodically poled potassium-titanyl-phosphate waveguide (PPKTP-WG) chip is pumped by a 440-nm pulsed laser, generating photon pairs with different polarization. The photons are then separated by a polarizing beamsplitter (PBS) after being filtered by an interference filter (IF) and two etalons, to a final linewidth of 1 GHz. Two half-wave plates (HWPs) oriented at \ang{22.5} are placed in each path. By implementing a two-photon interference on a second PBS, the photon pairs are prepared in the Bell state $\vert\Phi^{+}\rangle$ with post-selection. The phase shifts between polarizations introduced by optical components are compensated by wedge plates. One photon from each entangled photon pair is then sent to the middle for BSM through a 5-m fibre, and the other is stored in the quantum memory (QM). The photon in the BSM meets another photon from the other node with a time delay carefully synchronized by a right-angled prism and undergoes two-photon interference, resulting in four output channels, labelled ${\rm T}_{1}$, ${\rm R}_{1}$, ${\rm T}_{2}$ and ${\rm R}_{2}$. Clicks from the coincidence counting between ${\rm T}_{1}$ and ${\rm R}_{2}$ or ${\rm R}_{1}$ and ${\rm T}_{2}$ determine the $\vert\Phi^{+}\rangle$ state for the two photons, and thus heralding the state of the other two photons stored in the quantum memories to be $\vert\Phi^{+}_{M}\rangle$. The two quantum memories are placed in different cryostats separated by 3.5 m. The beam of single photons and a beam of pump light used for the preparation of the atomic frequency comb (AFC) are superimposed on the quantum memory with a small angle to reduce the noise. After a storage time of 55.6 ns, photons are released for analysis with a quarter-wave plate (QWP), a HWP and a PBS in each node. In our experiment, all of the single photons are collected by fibre collimators (FCs) and sent to an eight-channel superconducting nanowire single-photon detector (SNSPD) with a detection efficiency of about 85\% at a dark count rate of 10 Hz.
}\label{setup}
\end{figure}
\clearpage

\begin{figure}[tb]
\centering
\includegraphics[width= 0.7 \columnwidth]{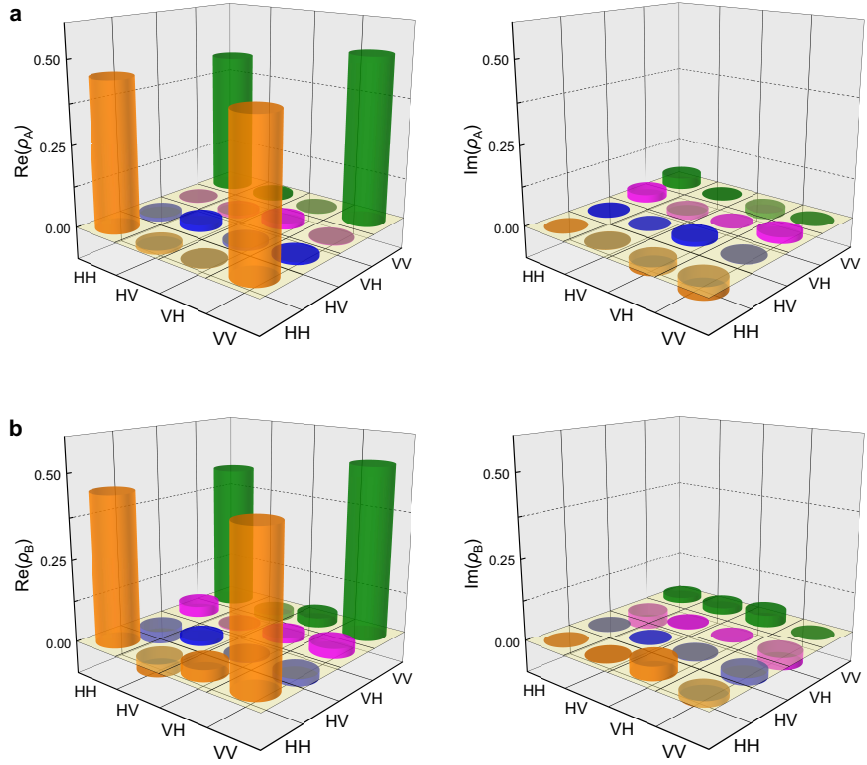}
\caption*{\justifying \textbf{Fig. 3: Quantum state tomography of two-photon polarization entanglement of two sources.} \textbf{a, b}, The reconstructed density matrix of entanglement source in Node A ($S_{A}$) and Node B ($S_{B}$) before storage of photons, denoted by $\rho_{A}$ and $\rho_{B}$, respectively. In the measurements, $g_{12}^{(2)}(\Delta\tau)$ is set to approximately 50 in a 2-ns detection window for both sources by choosing an appropriate pump power. The left figures represent their real part, Re($\rho_{A}$) and Re($\rho_{B}$), and the right figures represent their imaginary part, Im($\rho_{A}$) and Im($\rho_{B}$). 
}\label{sources}
\end{figure}
\clearpage

\begin{figure}[tb]
\centering
\includegraphics[width= 0.7 \columnwidth]{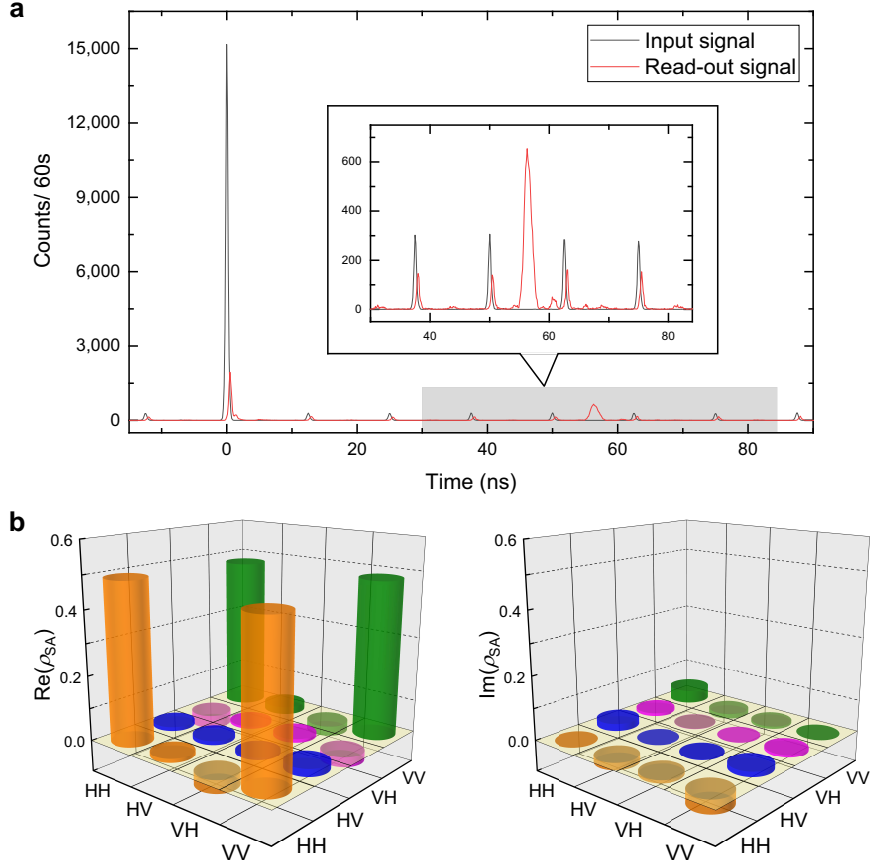}
\caption*{\justifying \textbf{Fig. 4: The performance of the quantum memory in node A.} \textbf{a}, Coincidence counts between photon pairs emitted from $S_{A}$ before and after storage by the quantum memory in node A. The bin width is 128 ps and the integration time is 60 s. The black line is the input signal and the red line is the read-out signal after a storage time of 55.6 ns. The inset shows an enlarged view of the indicated area in the histograms. The red peak in the middle represents the retrieved photons.  \textbf{b}, Tomographic results of polarization entanglement of $S_{A}$ after storage of photons in node A. The reconstructed density matrix is denoted by $\rho_{SA}$. The left figure represents its real part, Re($\rho_{SA}$), and the right figure represents its imaginary part, Im($\rho_{SA}$).
}\label{QM}
\end{figure}
\clearpage

\begin{figure}[tb]
\centering
\includegraphics[width= 0.7 \columnwidth]{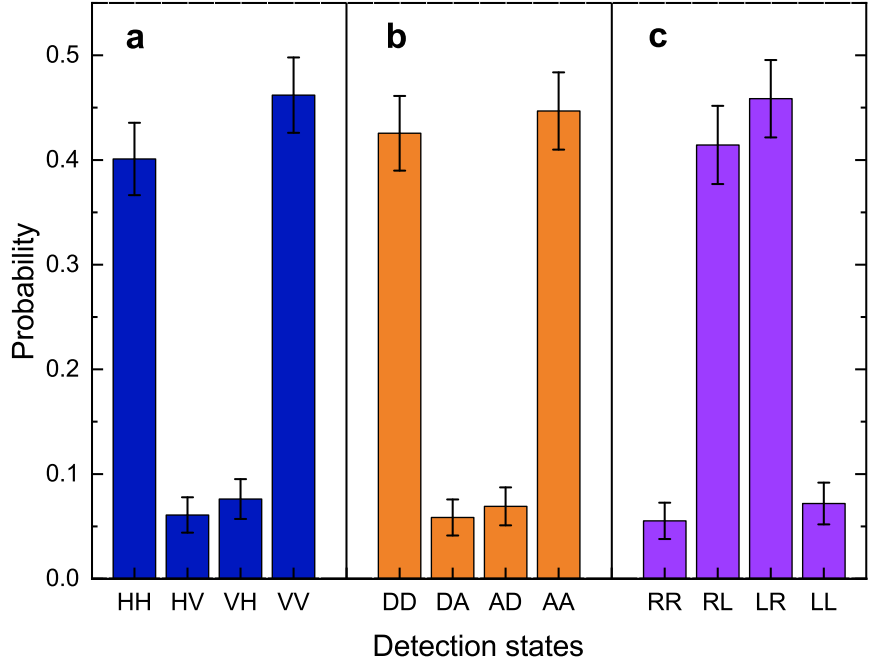}
\caption*{\justifying \textbf{Fig. 5: Verification of heralded remote entanglement between two quantum memories by polarization analysis of photons retrieved from two memories.} Three complementary bases are used according to the definition of quantum witness: \textbf{a}, H/V; \textbf{b}, D/A; \textbf{c}, R/L. The polarization bases are also chosen corresponding to the three different local measurements $\hat{\sigma}_{z}\hat{\sigma}_{z}$, $\hat{\sigma}_{x}\hat{\sigma}_{x}$ and $\hat{\sigma}_{y}\hat{\sigma}_{y}$ used to determine fidelity of the entanglement. Each measurement settings takes 160 h and the total measurement time is within two months. The bars represent measured correlation probabilities for different polarization states. The error bars are one standard deviation deduced from the Poisson statistics of the raw detection events based on Monte Carlo simulations.
}\label{MAIN}
\end{figure}
\clearpage

\setcounter{enumiv}{\value{firstbib}}
\setcounter{NAT@ctr}{\value{firstbib}}

\section*{\large Methods}
\textbf{Quantum memory samples.} 
We use the optical transition $^{4}I_{9/2}(0) \rightarrow {}^{4}F_{3/2}(0)$ of Nd$^{3+}$:YVO$_{4}$ crystals to store the polarization-entangled photons. The dimensions of crystals along $a\times b\times c$ are $3\times 8\times 14$ mm, respectively. The crystal, with light propagating along the $a$ axis, exhibits strong absorption for H polarization (parallel to the c axis) and much weaker absorption for V polarization. Thus, to realize nearly identical absorption depth for arbitrary polarization, each memory sample is made into a sandwich-like structure, which composed of two pieces of Nd$^{3+}$:YVO$_{4}$ crystals with 5-ppm doping concentration and a \ang{45} HWP sandwiched between them. Two memory samples are cooled to 1.5 K in separated closed-cycle cryostats (Oxford Instruments, SpectromagPT) integrated with a superconducting magnet. The magnetic field is set to 0.3 T parallel to the $c$ axis of crystals. Under this condition, the inhomogeneous broadening of optical transition is measured to be approximately 2.1 GHz, supporting efficient wideband storage.

\textbf{Timing sequence of the experiment.} 
The timing sequence of two nodes is synchronized by two arbitrary function generators (Tektronix, AFG3252). For each node, the preparation of AFC takes 3.8 ms. To isolate the fluorescence noise induced by the bright pump light, the storage and detection cycle start after a 0.6-ms wait time when the AFC preparation is completed. To protect the SNSPD from exposure to classic light and to further remove the unwanted noise, we place two phase-locked mechanical choppers in the pumping optical path and before the detection path in each node, respectively. Entangled photons are continuously stored in the quantum memories and retrieved for detection in a 5-ms cycle. The signals from SNSPD are sent to a logic module and a time-to-digital converter (Swabian, Time Tagger 20) for analysis. When the storage and detection cycle is finished, there is another 0.6-ms wait time and after that the AFC preparation cycle starts again. The complete experiment cycle is repeated at a fixed frequency of 100 Hz.

\textbf{Details about entangled photon-pair sources.}
Two identical polarization-entangled photon-pair sources are constructed based on the SPDC process. The detailed experimental setup is shown in Extended Data Fig. 1a. A pulsed Ti:sapphire laser (Spectra-Physics, Tsunami) at 880 nm with a pulse width of 300 ps and a repetition rate of 80 MHz pumps a PPKTP bulk crystal, producing 440-nm laser by second harmonic generation. The 440-nm laser is used to pump a PPKTP-WG chip (AdvR). The dimensions of the waveguide are 4 $\mu$m$\times$4 $\mu$m$\times$20 mm. An aspheric lens with $f= 8$ mm is used to focus the pump laser, achieving a total in-out efficiency of over $50\%$ including the coupling efficiency and transmission loss. Pairs of photons are generated by a type-II SPDC process. The photons are then filtered by a 1-nm bandpass interference filter and two etalons. The first etalon has a free spectral range of 220 GHz and a linewidth of 10 GHz and the second etalon has a free spectral range of 26 GHz and a linewidth of 1 GHz. These two etalons are both mounted in thermostat covers controlled by heating resistors and the temperature is stabilized at a specific point with a fluctuation of \SI{0.01}{\degreeCelsius}. The transmission efficiency of these two etalons at central frequency is approximately 95\% and 80\%, respectively. The design of this combination ensures a single longitudinal mode for filtered photon pairs with a linewidth of approximately 1 GHz. To generate polarization entanglement states, the photon pairs are separated by a PBS and then combined on a second PBS after going through a \ang{22.5} HWP in each path for two-photon interference \cite{postentanglement}.

The quantum correlation between photon pairs is limited by multi-pair emissions as a result of large emission probability when high pump power is used, which can be assessed by the second-order cross-correlation function $g_{12}^{(2)}(\Delta\tau)=P_{12}/(P_{1}P_{2})$ between two photons at zero time difference. Here, $P_{12}$ denotes the probability of detecting a coincidence between two photons within a time window of $\Delta\tau$, and $P_{1}$ ($P_{2}$) corresponds to the probability of detecting the the individual photons, where the subscripts 1 and 2 represent two output paths of the second PBS. The cross-correlation can be directly measured in the experiment through the coincidence counting between the photons. Extended Data Fig. 1b, c shows the counting rate and $g_{12}^{(2)}(\Delta\tau)$ as a function of the pump power for the two sources, respectively. The coincidence window is chosen as $\Delta\tau=2$ ns, which would comprise nearly all the photons in the temporal profile. The coincidence counting rates, which stand for the brightness of the source, grow linearly with pump power. Nevertheless, $g_{12}^{(2)}(\Delta\tau)$ exhibits quadratic decrease to the pump power. To reduce the effect of multi-pair emissions while maintaining an acceptable photon-counting rate, we choose a balanced value of $g_{12}^{(2)}({2\, \rm ns})=50(4)$ for both sources in the main experiment to generate heralded entanglement between two quantum memories. In this case, neglecting multi-pair emissions, the state of generated photon pairs after the second PBS becomes
\begin{equation}\label{S12}
\begin{aligned}
\vert\psi\rangle_{12}=\dfrac{1}{2} \left(  \vert H\rangle_{1}\vert H\rangle_{2} + \vert V\rangle_{1}\vert V\rangle_{2}- i\vert HV\rangle_{1}\vert {\rm vac}\rangle_{2} +i\vert {\rm vac}\rangle_{1}\vert HV\rangle_{2} \right),
\end{aligned}
\end{equation}

where the subscripts 1 and 2 represent two output paths and $\vert {\rm vac}\rangle$ is the vacuum state with no photon in this path. Photons in path 1 are sent to quantum memory while photons in path 2 are sent to the middle for BSM. The purity of the sources is quantified by quantum state tomography. In each path, we place a qubit polarization analyser that contains a QWP, a HWP and a PBS. Photon pairs are measured with 16 sets of selected polarization combination to determine the corresponding two-photon correlations. The density matrix of entangled photon pairs is reconstruct based on these measurements using maximum-likelihood estimation \cite{tomo}. It is worth noting that the characterization of the entangled photon-pair sources with quantum state tomography requires measuring the coincidence counts of two paths, which post-select a maximally entangled polarization state while inducing 50\% photon loss. Considering the case of one photon in each output paths, the state of photon pairs can be represented as
\begin{equation} 
\vert\Phi^{+}\rangle_{12}=\dfrac{1}{\sqrt{2}}\left(\vert H\rangle_{1}\vert H\rangle_{2} + \vert V\rangle_{1}\vert V\rangle_{2}\right).
\end{equation}
The fidelities are 92.6(3)\% and 93.3(3)\% for the two sources with $g_{12}^{(2)}({2\, \rm ns})=50(4)$ (see Fig. 3), respectively, confirming the generation of high-purity entanglement.

The coincidence counting rate of the photon pairs generated directly from PPKTP-WG at $g_{12}^{(2)}({2\, \rm ns})=50(4)$ is around $1.25\times10^{4}$ s$^{-1}$, with an average single channel counts of approximately $1.5\times10^{5}$ s$^{-1}$. The average heralding efficiency is 8.3\%. After post-selection of two-photon entanglement, the counting rates of two entangled photon-pair sources become approximately 5,700 s$^{-1}$ and 6,200 s$^{-1}$ for node A and node B, respectively.

\textbf{Bell-state measurement.}
The BSM used in the experiment consists of a type-\uppercase\expandafter{\romannumeral2} fusion nondeterministic gate \cite{type2fusion} (realized by placing one HWP in each of the two inputs and two outputs of a PBS) and subsequent polarization-selective photon detections. A perfect BSM requires the photons from the different sources (labelled as photons 2 and 3) arrive at the first PBS at the same time. This condition is ensured by carefully tuning the length of one optical path with a prism.
 
For ideal entanglement sources, the BSM can distinguish between the two Bell states,  $\vert\Phi^{+}\rangle_{23}$ and $\vert\Psi^{+}\rangle_{23}$, with a probability of 1/2 by different click patterns of four detection ports, labelled as ${\rm T}_{1}$, ${\rm R}_{1}$, ${\rm T}_{2}$ and ${\rm R}_{2}$. In practice, with the presence of two-photon terms ($\vert HV\rangle_{2}$ and $\vert HV\rangle_{3}$), the success probability of BSM becomes 1/8. To see this, we can write the state of the whole photonic system before BSM as
\begin{equation}\label{SS}
\begin{aligned}
\vert\psi\rangle_{12}\otimes\vert\psi\rangle_{34}=&\dfrac{1}{4} \left(  \vert H\rangle_{1}\vert H\rangle_{2} + \vert V\rangle_{1}\vert V\rangle_{2}-i\vert HV\rangle_{1}\vert {\rm vac}\rangle_{2} +i\vert {\rm vac}\rangle_{1}\vert HV\rangle_{2} \right)\otimes\\
& \left(  \vert H\rangle_{3}\vert H\rangle_{4} + \vert V\rangle_{3}\vert V\rangle_{4}-i \vert HV\rangle_{3}\vert {\rm vac}\rangle_{4} +i\vert {\rm vac}\rangle_{3}\vert HV\rangle_{4} \right).
\end{aligned}
\end{equation}
In the Schmidt decomposition of equation (7), only four of the total 16 terms are useful for successful entanglement swapping when there are photons in both 1, 2, 3 and 4. The remainder are unwanted but can be eliminated by projective measurement during BSM. After passing \ang{22.5} HWPs before the first PBS in the BSM, two-photon terms $\vert HV\rangle$ ($\vert HV\rangle_{2}$ and $\vert HV\rangle_{3}$) are transformed to $\left(1/\sqrt{2} \right) \left( \vert HH\rangle-\vert VV\rangle \right)$, which means the two photons go to the same outputs of the first PBS. Thus, all seven terms containing two-photon terms do not have the contribution to the click patterns of successful BSMs, which require only one photon at each output of the first PBS. Five terms containing zero or only one photon in the BSM are insufficient to provide two clicks. Consequently, only four remaining terms can provide the correct and useful click patterns of photon detectors, yielding the success probability of BSM to be 1/8. Because of the imperfect experimental conditions, such as the presence of photon loss and the lack of photon-number-resolving detectors, unwanted terms can not be completely eliminated and thus add spurious events in the BSM. Fortunately, these events herald the states with at most one photon in both path 1 and 4 and would not affect the fidelity of distributed two-photon states. In our experiment, the spurious events can be excluded during the stage of fourfold coincidence measurement of photons 1, 2, 3 and 4 when verifying the heralded entanglement.

Considering effective terms during the fourfold coincidence measurement, the operations in BSM correspond to standard entanglement swapping \cite{sdswapping}. Note that the BSM occurs when photon 1 and 4 are stored in the quantum memories. Conditioned on registering coincidence counts between detectors of ${\rm T}_{1}$ and ${\rm R}_{2}$ ,or ${\rm R}_{1}$ and ${\rm T}_{2}$, $\vert\Phi^{+}\rangle_{23}$ is selected and the two remote quantum memories are projected onto a similar entangled state $\vert\Phi^{+}_{M}\rangle_{14}$, as explained in the main text.

It should be pointed out that the multi-pair emissions also add spurious events to the overall rate of BSMs that do not result in successful entanglement swapping \cite{twoes2}. Such events are fully excluded at the stage of fourfold coincidence measurement and can be washed out after subsequent entanglement connections with properly designed BSMs \cite{twophoton,twoes2}, thus do not hinder the applications of our experimental configuration in quantum repeaters. Nevertheless, we stress that probabilistic entanglement sources generated from the SPDC used in this work are for the proof-of-principle demonstration of an elementary link of a quantum repeater based on absorptive quantum memories. The absorptive quantum memories combined with deterministic entanglement sources  \cite{QDE4,QDE1,QDE2,QDE3} would be the best option for the construction of efficient quantum repeaters in the near-term future  \cite{semihiera}.

\textbf{Entanglement swapping with memory-compatible photons.}
The generation of independent entanglement photon sources with quantum-memory-compatible linewidth and demonstration of entanglement swapping between them is the first step towards quantum repeater architecture based on absorptive quantum memories. So far, very few experiments have explored this, except for the work demonstrated by Jin \emph{et al}. \cite{jinswapping}. They used the time-bin entangled photon-pair source, which is designed for Tm-doped waveguide quantum memories with a bandwidth of 6 GHz. However, the achieved final state fidelity is only 68(3)\% after entanglement swapping with an entanglement generation rate of 10 h$^{-1}$. Such low-fidelity and low-rate entanglement would not allow a violation of Bell's inequality and prevents any further demonstration with quantum memories. 

In our experiment, we use the polarization degree of freedom, which is easy to manipulate and does not need any additional actively locked interferometers to analyse the qubit. We perform the entanglement swapping experiment without the storage of photons. A click of the BSM heralds the generation of entangled two-photon state $\vert\Phi^{+}\rangle$ at two nodes. We vary the pump power and measure the fidelity of heralded photon pairs after entanglement swapping. Extended Data Fig. 2a shows measured fidelity for different $g_{12}^{(2)}({2\, \rm ns})$. The fidelity for $g_{12}^{(2)}({2\, \rm ns})=40$ is 73.1(1)\% with an average fourfold coincidence counting rate of 1,200 h$^{-1}$, which is 120 times enhanced than that reported in ref. \cite{jinswapping}.

To further quantify the quality of the heralded two-photon entanglement, we measure the power dependence of the $S$ parameter in a CHSH-type Bell's inequality \cite{CHSH}, which is presented as
\begin{equation} \label{S}
S=|E\left(a,b\right)+E\left(a,b'\right)+E\left(a',b\right)-E\left(a',b'\right)|.
\end{equation}
Here, $E\left(a,b\right)$ represents the correlation coefficient, and $a$ $(a')$ and $b$ $(b')$ are the polarization bases for the measurement of two heralded entangled photons, which is realized through the combination of a QWP, a HWP and a PBS. In the measurement, we set $a=\left(\ang{0},\ang{0}\right)$, $a'=\left(\ang{45},\ang{22.5}\right)$, $b=\left(\ang{22.5},\ang{11.25}\right)$, and $b'=\left(\ang{67.5},\ang{78.75}\right)$, where the two angles correspond to the orientation of the QWP and HWP, respectively. The $S$ parameter as a function of $g_{12}^{(2)}({2\, \rm ns})$ is shown in Extended Data Fig. 2b. The highest $S$ parameter we obtained is $S=2.427\pm0.13$ for $g_{12}^{(2)}({2\, \rm ns})=113$, which violates Bell's inequality by more than 3 standard deviations. 

For the demonstration of heralded entanglement between two quantum memories, we choose $g_{12}^{(2)}({2\, \rm ns})=50(4)$ for both sources. In this case, the fourfold coincidence counting rate of heralded photon pairs before storage is 410 h$^{-1}$. The storage efficiencies of quantum memories in the two nodes are 14.3(1)\% and 12.5(1)\%, respectively, at a storage time of 55.6 ns. The end-to-end efficiencies of memories are reduced to 9.7\% and 6.4\% due to the transmission losses, window reflections and coupling losses. Considering 50\% duty cycle of the memory process, the expected fourfold coincidence counting rate after storage is $410\times 9.7\% \times 6.4\% \times 50\%=1.27$ h$^{-1}$. This is consistent with the experimental achieved rate of 1.1 h$^{-1}$, proving the stability of our setup over the whole measurement time of two months. The fidelity of the heralded entangled photon pairs before storage is 76.9(2)\%. After storage, the fidelity is measured to be 80.4(2)\%. The slight increase in fidelity is caused by high-quality filtering of the quantum memory process. Similar phenomenon has been observed in many previous publications  \cite{QME1,QME2,mode1}. 

\textbf{Preparation of wideband AFC.}
The pump light used in AFC preparation is generated from a Ti:Sapphire laser (MBR-110, coherent) with a frequency of 340.6967 THz. The frequency of the laser is stabilized to a low-drift Fabry-P\'erot Interferometer placed in a high-vacuum housing to achieve a linewidth of approximately 100 kHz. We use an acousto-optic modulator (AOM) in a double-pass configuration to control the frequency and amplitude of the AFC pump light. The frequency of the AOM is swept over 198 MHz repeatedly during the preparation process. Each frequency point is assigned a specific amplitude, which is optimized by storage efficiency to give an optimal comb structure. The period of the AFC is 18 MHz, corresponding to a storage time of 55.6 ns. The bandwidth of AFC is further extended to approximately 1 GHz using a fibre-pigtailed electro-optic phase modulator (EO-PM, Eospace) \cite{mode1}. We use an RF signal generator with a frequency of 198 MHz and tunable amplitude to create frequency sidebands of the pump laser. All four sidebands ($\pm 2$ and $\pm 1$ sidebands) are generated simultaneously while the frequency of the AOM is swept to create five 198-MHz AFCs in parallel, forming the total AFC structure with a bandwidth up to 1 GHz. An example of a measured wideband AFC structure with 1-GHz bandwidth is shown in Extended Data Fig. 3. The modulation depth of the EO-PM is carefully optimized according to the storage efficiency. 

The pump beam and signal beam are overlapped at the sample with a non co-linear configuration. The signal beam is focused to a diameter of 90 $\mu$m, while the pump beam is expanded to produce a larger beam waist ($\sim$ 250 $\mu$m) on the sample to ensure a good overlap with the signal beam.

\textbf{Memory time and storage efficiency.}
Normally, the storage efficiency is associated with memory time due to decoherence of the optical excitations. To study the memory capability in our memory samples, we measure the intensity of retrieved AFC echo as a function of storage time. The results are shown in Extended Data Fig. 4. We fit the data with a double exponential decay and obtain $\tau_{1}=134$ ns and $\tau_{2}=1,141$ ns, much smaller than the optical coherence time measured through photon echo (approximately 60 $\mu$s). Except for technical limitations, such as the frequency noise of the laser and the vibration of the cryostat, the two decay lifetimes may be caused by the presence of even and odd isotopes of Nd$^{3+}$ and the superhyperfine interaction between the electron spins of Nd$^{3+}$ and the nuclear spins of Y$^{3+}$ and V$^{3+}$ in the host crystal.

It is worth mentioning that for photon-echo-based quantum memories such as AFC protocol, the decoherence only affects the efficiency of the memory, but not the fidelity of the output state, as long as the background noise is negligible. This can be understood that the decohered ions would not contribute to the collective interference that produces the AFC echo. Two phase-locked mechanical choppers are employed to completely gate the AFC pump light. To avoid fluorescence, we set the time between memory process and AFC pump process to 0.6 ms, which is much greater than the optical population lifetime. Therefore, in our setup, the high memory fidelity could be maintained for longer storage times \citep{highF}, with the sacrifice of storage efficiency.

For practical long-distance quantum repeaters, a storage time above milliseconds using a spin-wave-based storage protocol is required. Currently, this can be only achieved with non-Kramers ions, such as Eu$^{3+}$, and unfortunately, the bandwidth is typically limited to a few MHz \cite{euspinwave}. To achieve wideband spin-wave storage, one could utilize hybrid electron-nuclear spins of odd isotopes of Nd ($^{143}$Nd, $^{145}$Nd) which have nuclear spins of 7/2. Zero-first-order-Zeeman (ZEFOZ) magnetic fields will be required to minimize its transition sensitivity to magnetic noises inside the host crystal and thus suppress the superhyperfine interaction \cite{Ybnm,superhyperfine}. The most challenging part in implementing spin-wave storage in Nd$^{3+}$-doped material is the complicated and unknown level structure for the odd isotopes of Nd ions. Our recent work has completely identified the ground-state level structure of $^{143}$Nd at sub-kelvin temperatures \cite{NdEPR}. The next thing to do is to sort out the excited-state hyperfine-level structure and to identify a proper $\Lambda$-system for optical spin-wave storage. Other Kramers ions such as Er$^{3+}$ (ref. \citep{Erkramers}) and Yb$^{3+}$ (ref. \cite{Ybnm}) can also be candidates, and the latter has already enabled the spin-wave-AFC storage of classical light for over 1 ms (ref. \cite{YbPRL}), which indicates that wideband and long-lived quantum storage should be feasible using Kramers ions.

As a wideband quantum memory based on Nd-ion-doped crystal, the storage efficiency is approximately three times of that reported with Nd$^{3+}$:Y$_{2}$SiO$_{5}$ crystal \cite{Ndeff}. The overall storage performance (efficiency$\times$time$\times$bandwidth) is 7.8. To our knowledge, this is the best value reported so far for a wideband solid-state quantum memory \cite{band16g,widebandeff2}. Such enhancement is crucial to establish the two-photon entanglement in our approach since the final successful probability depends on the square of memory efficiency.

\textbf{Temporal multiplexed operations and entanglement distribution rate.}
The final fourfold photon counting rate--that is, the entanglement distribution rate (EDR)--achieved in the experiment is 1.1 h$^{-1}$. After the calibration of the loss after the quantum memories, including the recall efficiency (defined as the storage efficiency divided by the absorption probability, $\sim$16\% for each node), the transmission along the optical path ($\sim$75\% for each node) and the detection efficiency ($\sim$85\% for each channel), the rate of heralded entanglement stored in the memories is approximately 106 h$^{-1}$.

In the quantum repeater protocol, for each attempt of entanglement distribution, one elementary link must wait the time required for the photons to transmit from the sources to the intermediate station and for the result of BSM to send back to the quantum memories. The waiting time $L_{0}/c$, where $L_{0}$ is the channel distance of two remote memories and $c$ is the speed of light, is called communication time, which should be shorter than memory storage time. The use of multimode quantum memory can greatly improve the EDR \cite{multiQP} since more than one mode are simultaneously stored in the quantum memory at each attempt and it is not necessary to wait for the heralding signal from the BSM before storing another mode.

Experimentally, we utilize the temporally multiplexed operation to enhance the final EDR. We use a pulsed laser to pump the nonlinear crystals, with a repetition of 80 MHz. The separation of pump pulses is 12.5 ns and the time width of photon pairs is about 1 ns (1-GHz linewidth). Thus, photon pairs generated in different pulses can be treated as distinguished temporal modes. By virtue of the temporal multimodality of the AFC storage \citep{AFC}, four temporal modes can be stored in the quantum memory within the storage time before the first absorbed mode has been released, see Extended Data Fig. 5a. The mode can be identified and sorted by the synchronization signal from the pulsed laser. Since only a single elementary link is demonstrated in the current work, these modes are not sorted and are joined together to accelerate the data acquisition. The maximum number of temporal modes available is approximately equal to the storage time divided by the time width of the stored photons, that is, the time-bandwidth product (TBP) \cite{TBP,AFC}. In our configuration, the TBP is about 56, which means that the maximum supported mode number is 56. However, the repetition rate of the pump laser restricts the mode number realized in the experiment to about 55.6 ns/12.5 ns $\approx$ 4. The entanglement distribution can increase linearly with the number of modes. Therefore, the four modes used in our experiment help directly to enhance the EDR rates by 4 times. If we fully utilize the 56 modes with high-repetition laser, the EDR could be boosted to approximately 15.4 h$^{-1}$. The estimated EDR with respect to the increase of mode number is shown in Extended Data Fig. 5b.

The current EDR is primarily limited by the probabilistic generation of entangled photon-pairs through the SPDC process. The production probability of entangled photon-pair is approximately 0.012 for a single pump pulse (including 50\% photon loss in the post-selection of two-photon entanglement). The four-photon generation probability is therefore $1.44\times10^{-4}$. The heralding efficiency of the SPDC source achieved in this work is approximately 8\%, corresponding to a pair collection efficiency of 0.64\% for entangled photon-pairs. By updating the photon sources to perfect deterministic entangled photon sources which in principle allow a pair emission probability of 100\%, the EDR would be enhanced by $1.7\times10^{8}$ times. The deterministic photon sources will increasingly accelerate the EDR when going to multiple repeater nodes. The efficient interface with deterministic source  \cite{mode2} is the fundamental motivation for our choice of absorptive wideband quantum memories.

Except for the photon sources, the storage efficiency of quantum memory used in this work also needs to be improved before practical applications. Using cavity enhancement technique, the memory efficiency can be in principle enhanced to close to unity \citep{cavity1}. The highest quantum storage efficiency demonstrated in rare-earth-ion-doped crystals so far is 69\% \cite{higheff}. If this storage efficiency is attained in our demonstration, the total system efficiency and final EDR could be enhanced by approximately 26 times. Considering the use of the state-of-the-art deterministic entangled photon-pair source at the same time \cite{QDE2}, which has the pair emission rate of approximately $7.3 \times 10^{5}$ s$^{-1}$ before single-photon detectors with a repetition rate of 76 MHz, the EDR could reach approximately $9 \times 10^{5}$ h$^{-1}$. If the repetition rate of the source is boosted to 1 GHz to take full advantage of temporally multiplexed operations, the EDR could further be enhanced to $1.2 \times 10^{7}$ h$^{-1}$, over seven orders of magnitude greater than the current EDR. This shows the great potential of our system, which is especially suitable for the construction of high-speed quantum repeaters and quantum networks in the near future.

\textbf{Fidelity of heralded remote entanglement between two quantum memories.}
The fidelity of final entanglement between quantum memories is mainly limited by multi-pair emissions. The effect of multi-pair emissions has been mentioned above. There is a fundamental trade-off between fidelity and generation probability for probabilistic photon source such as the SPDC source. When lowering the pump power, the fidelity of the source can be increased at the expense of low counting rate.

In the BSM, we use two-photon interference, which does not require phase stability. The laser and propagation phases only contribute to an irrelevant global phase. The two-photon interference does need the temporal overlap of wavepacket between the two photons, which depends on the coherence length of photons. In our experiment, the coherence length of photons is on the order of hundreds of mm, which does not need any stabilization. Thus, the phase noise is negligible in our demonstration.

The background noise, which may decrease the visibility of the entangled photon pairs, is negligibly small as a consequence of fourfold photon coincidence counting. The noise of the AFC pump is eliminated in the setup, as described above. After carefully covering all the fibres and optical paths, we obtain an average background noise of about 200 Hz including the dark counts of the SNSPD. Thus, the experimental noise of coincidence counting between two retrieved photons after quantum memories (within a detection windows of 3 ns) can be estimated to be $(200 \, {\rm s}^{-1}\times3 \, {\rm ns})^{2}=3.6\times10^{-13}$ per heralding signal. The heralding rate of BSM is approximately 100 s$^{-1}$. Thus, the final experimental noise of fourfold coincidence counting is  $3.6\times10^{-13}\times 100 \, {\rm s}^{-1}\times3,600 \, {\rm s \, h^{-1}}=1.3\times10^{-7}$ h$^{-1}$, far below the signal of 1.1 h$^{-1}$ obtained in the experiment.

Furthermore, for protocols based on two-photon detection in the BSM, the photon loss only affects the rate of the experiment, but not the fidelity of the distributed entanglement when the background noise is negligible. Thus, imperfect heralding efficiency and memory efficiency do not lead to the decrease of the fidelity. However, if the memory efficiency decreases to a value when the coincidence counts of retrieved photons are comparable to the background noise, the fidelity of entanglement would begin to noticeably decrease. To explore the limit of memory efficiency, we calculate the estimated fidelity of heralded remote entanglement between two quantum memories as a function of storage efficiency, based on the fidelity and noise measured in the experiment. The results are plotted in Extended Data Fig. 6. As can be seen from the figure, the fidelity of distributed entanglement is particularly robust to the storage efficiency. The quantum memories could still provide a useful storage when the storage efficiency drops to approximately 0.01\% where the retrieved two-photon entanglement still has a fidelity higher than classical bound of 0.5, although a direct characterization of such entanglement is not practical because of the low efficiencies and long integration times.

\textbf{Heralding signal and heralding probability.} 
The heralding feature is important to implement quantum repeater protocol, especially when synchronizing different elementary links. In theory, the heralding signal should send to the quantum memories before the photons are released from quantum memories, in other words, memory time should be longer than the communication time \cite{reviewQP}. Our setup meets the requirement of heralding despite a short heralding time. Here, we provide a detailed analysis of the system delay in our experiment. Since no actual heralding for synchronization process is implemented in this single-link experiment, we only provide all related delays and analyse the allowed heralding time according to all known delays. In our setup, one of the photons from an entangled photon-pair in each node is sent to the middle station for BSM through a 5-m fibre, and the other photon is sent to a quantum memory through a 3-m fibre for storage. The photon arrived at the middle station after the other photon is captured by the quantum memory for $(5-3)/(2\times10^{8}) = 10$ ns. Here we neglect the free-space propagation delays since they are approximately the same in the memory station and the middle station. The photons in the BSM are detected by SNSPD through 1-m fibres (5-ns delay). The delay for single-photon detection, as provided by the manufacturer for the SNSPD, is typically 15 ns, taking account of optical and electrical delays. The propagation delay of the logic module (CO4020, ORTEC) is measured to be 8 ns. The distance between the BSM and the memories is 1.8 m. The heralding signal should be sent back to the quantum memories, which can be directly transmitted through free space, inducing a delay of $1.8/(3\times10^{8}) = 6$ ns. Therefore, the heralded entanglement is available for about $(55.6-10-5-15-8-6) \approx 11.6$ ns. However, note that the quantum memory used in this work holds an efficiency over 1\% for a storage time of 1,250 ns (see Extended Data Fig. 4), which provides a useful storage where the retrieved two-photon entanglement still has a fidelity considerably higher than the classical bound, according to the simulation shown in Extended Data Fig. 6. This would provide a heralding time much longer than the communication time.

The final fourfold coincidence rate after SNSPD achieved in our demonstration is approximately 1.1 h$^{-1}$, with an average heralding rate of 100 Hz. Therefore, the heralding probability of BSM is $1.1/(100\times3,600)=3.06\times10^{-6}$. The low heralding probability is due to fourfold coincidence counting, which is mainly limited by the memory efficiency and heralding efficiency of the SPDC source. In principle, using single-photon detection in the BSM to generate number-state entanglement between quantum memories can achieve a higher heralding rate. For example, if we adopt the single-photon-based approach, the counting rate in a single channel of each SPDC source is approximately 150 kHz in our experiment. Then the expected heralding rate is 150 kHz for generation of the single-photon entanglement, as measured at one output of the beamsplitter in the BSM. The heralding rate is more than three orders of magnitude than that in two-photon-detection approach. However, this has a price since single-photon interference requires strict phase stability of the pump lasers and the fibre lengths, which is extremely difficult for practical long-distance applications. In addition, to remove the large vacuum component and generate useful two-photon entanglement, the implementation of two chains of number-state entanglement between quantum memories in parallel is finally required.

~\\

\noindent{\bf  Data availability}
The data presented in the figures within this paper and other findings of this study are available from the corresponding authors upon reasonable request.

\noindent{\bf  Code availability}
The custom codes used to produce the results presented in this paper are available from the corresponding authors upon reasonable request.

\clearpage

\begin{figure}[tb]
\centering
\includegraphics[width= 1 \columnwidth]{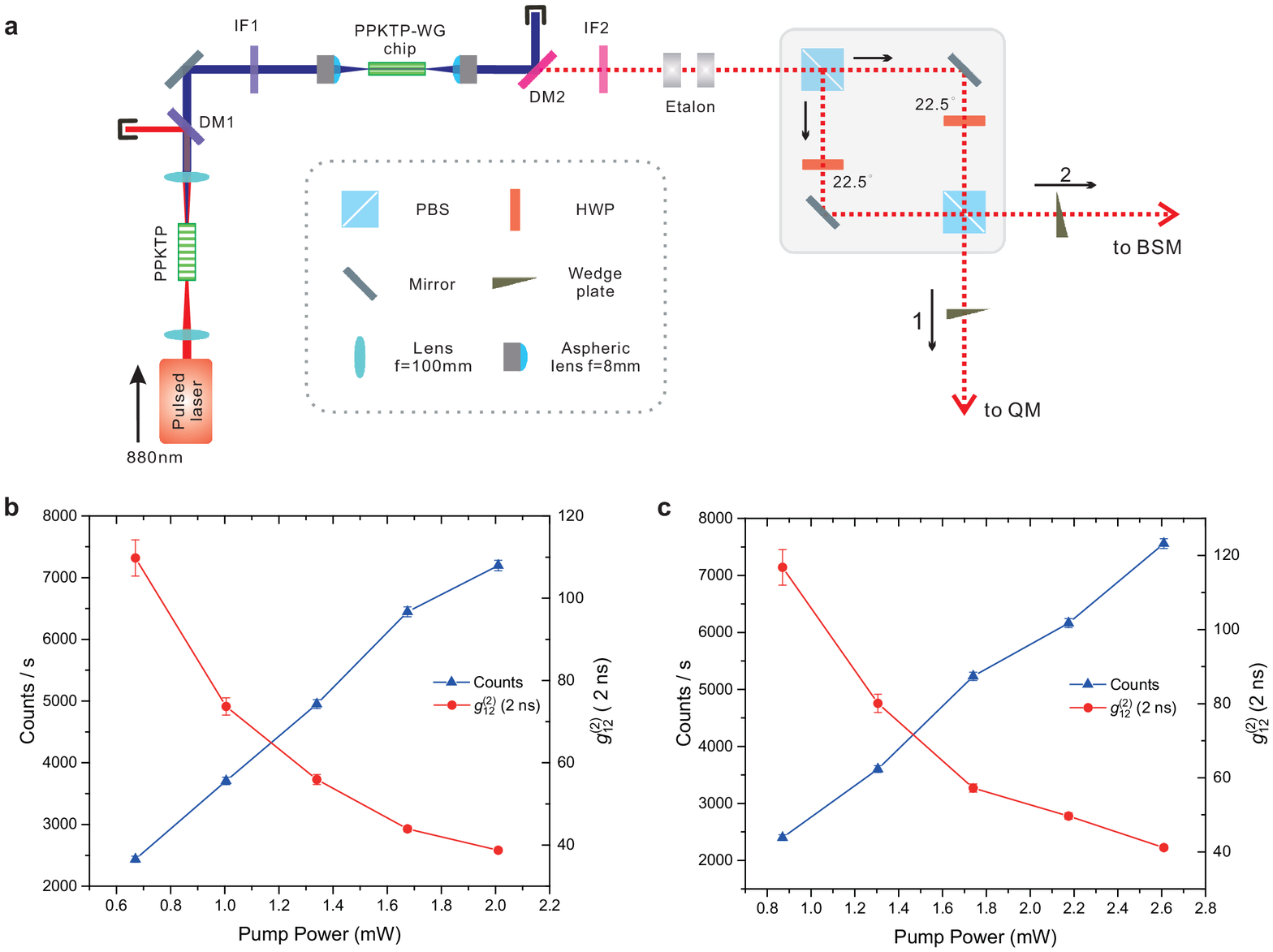}
\caption*{\justifying \textbf{Extended Data Fig. 1: Experimental setup and characterizations of entangled photon-pair sources.} \textbf{a}, Detailed setup for entangled photon-pair sources. \textbf{b, c}, Power dependence of the coincidence counting rate and second-order-correlation function $g_{12}^{(2)}({2\, \rm ns})$ for each entangled photon-pair source located at two nodes, respectively. Error bars in \textbf{b} and \textbf{c} represent one standard deviation. DM, dichroic mirror; IF, interference filter; QM, quantum memory.
}
\end{figure}

\clearpage

\begin{figure}[tb]
\centering
\includegraphics[width= 1 \columnwidth]{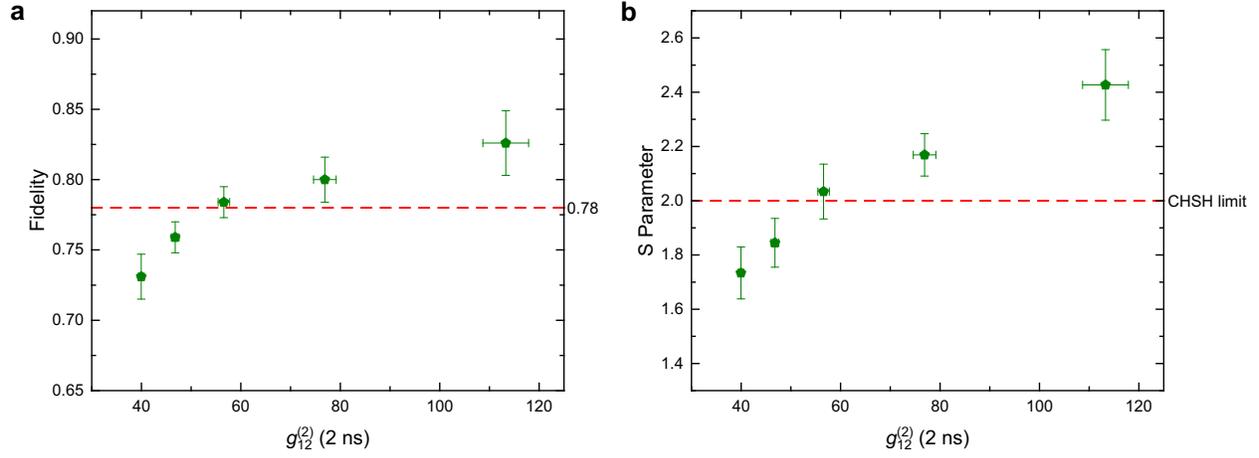}
\caption*{\justifying \textbf{Extended Data Fig. 2: Characterization of the heralded entangled photon pairs.} \textbf{a}, The fidelity of heralded entangled photon pairs as a function of average $g_{12}^{(2)}({2\, \rm ns})$ of two sources. \textbf{b}, Measured $S$ parameter of CHSH-type Bell's inequality with respect to the average $g_{12}^{(2)}({2\, \rm ns})$. Error bars are one standard deviation.
}
\end{figure}

\clearpage

\begin{figure}[tb]
\centering
\includegraphics[width= 0.7 \columnwidth]{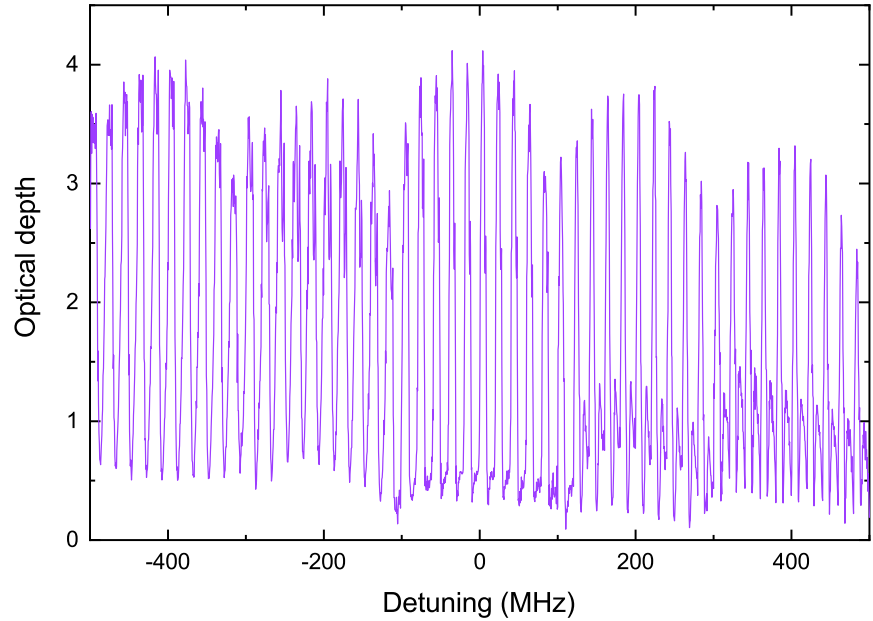}
\caption*{\justifying \textbf{Extended Data Fig. 3: An example of an AFC structure with a bandwidth of 1 GHz.} The central 200-MHz AFC is generated by an AOM, and the four sidebands are generated by an EO-PM in parallel. The total AFC bandwidth is approximately 1 GHz. The AFC structure is determined by the transmission of weak probe light using single-photon detectors. The polarization of the probe light is chosen as $H+V$.
}
\end{figure}

\clearpage

\begin{figure}[tb]
\centering
\includegraphics[width= 0.7 \columnwidth]{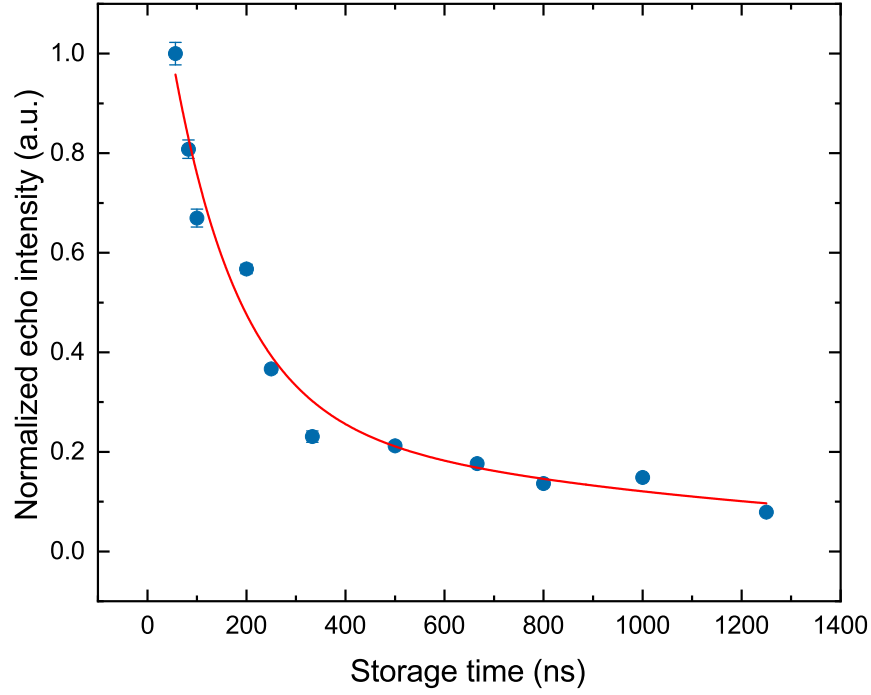}
\caption*{\justifying \textbf{Extended Data Fig. 4: The AFC echo intensity as a function of storage time.} The echo intensity (blue point) is normalized to the value of 55.6-ns storage time. Error bars represent one standard deviation. The red line is double exponential fit ($Ae^{-t/\tau_1}+Be^{-t/\tau_2}$) of the experimental data, with $A=1.04$, $B=0.29$, $\tau_{1}=134$ ns and $\tau_{2}=1,141$ ns. The $1/e$ lifetime of storage efficiency is deduced to be 193 ns based on fitted data. a.u., arbitrary units. 
}
\end{figure}

\clearpage

\begin{figure}[tb]
\centering
\includegraphics[width= 1 \columnwidth]{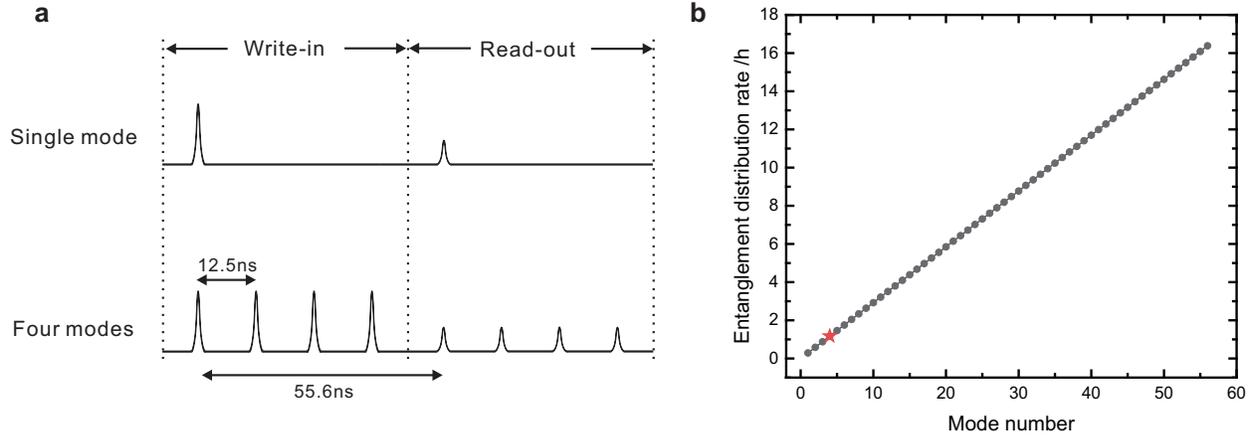}
\caption*{\justifying \textbf{Extended Data Fig. 5: Temporal multiplexed operations of a quantum memory.} \textbf{a}, A schematic diagram of temporally multiplexed operation. Compared to a single-mode scenario, four temporal modes are stored within the 55.6-ns storage time in the experiment. \textbf{b}, The estimated entanglement distribution rate (EDR) as a function of the mode number. The red star is achieved EDR with multiplexing of four time modes achieved in the experiment and the grey dots are estimated values based on experimental data.
}
\end{figure}

\clearpage

\begin{figure}[tb]
\centering
\includegraphics[width= 0.7 \columnwidth]{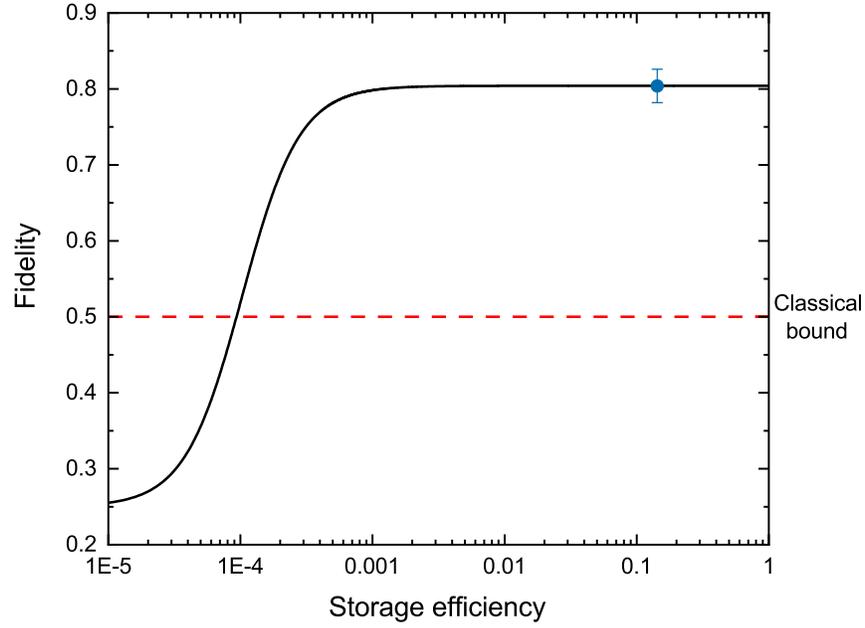}
\caption*{\justifying \textbf{Extended Data Fig. 6: Estimated fidelity of heralded remote entanglement between two quantum memories as a function of storage efficiency.} The blue dot is the data measured in the experiment and the black solid line is the simulation based on experimentally determined background noise. The red dashed line represents the fidelity of the classical bound. The error bar of the blue dot is one standard deviation.
}
\end{figure}



\begin{thebibliography}{10}
\expandafter\ifx\csname url\endcsname\relax
  \def\url#1{\texttt{#1}}\fi
\expandafter\ifx\csname urlprefix\endcsname\relax\def\urlprefix{URL }\fi
\providecommand{\bibinfo}[2]{#2}
\providecommand{\eprint}[2][]{\url{#2}}

\bibitem{ED100KM}
\bibinfo{author}{Yin, J.} \emph{et~al.}
\newblock \bibinfo{title}{Quantum teleportation and entanglement distribution
  over 100-kilometre free-space channels}.
\newblock \emph{\bibinfo{journal}{Nature}} \textbf{\bibinfo{volume}{488}},
  \bibinfo{pages}{185--188} (\bibinfo{year}{2012}).

\bibitem{firstQP}
\bibinfo{author}{Briegel, H.-J.}, \bibinfo{author}{D\"ur, W.},
  \bibinfo{author}{Cirac, J.~I.} \& \bibinfo{author}{Zoller, P.}
\newblock \bibinfo{title}{Quantum repeaters: The role of imperfect local
  operations in quantum communication}.
\newblock \emph{\bibinfo{journal}{Phys. Rev. Lett.}}
  \textbf{\bibinfo{volume}{81}}, \bibinfo{pages}{5932--5935}
  (\bibinfo{year}{1998}).

\bibitem{QNcold1}
\bibinfo{author}{Chou, C.-W.} \emph{et~al.}
\newblock \bibinfo{title}{Functional quantum nodes for entanglement
  distribution over scalable quantum networks}.
\newblock \emph{\bibinfo{journal}{Science}} \textbf{\bibinfo{volume}{316}},
  \bibinfo{pages}{1316--1320} (\bibinfo{year}{2007}).

\bibitem{QNcold2}
\bibinfo{author}{Yuan, Z.-S.} \emph{et~al.}
\newblock \bibinfo{title}{Experimental demonstration of a {BDCZ} quantum
  repeater node}.
\newblock \emph{\bibinfo{journal}{Nature}} \textbf{\bibinfo{volume}{454}},
  \bibinfo{pages}{1098--1101} (\bibinfo{year}{2008}).

\bibitem{QNcold4}
\bibinfo{author}{Yu, Y.} \emph{et~al.}
\newblock \bibinfo{title}{Entanglement of two quantum memories via fibres over
  dozens of kilometres}.
\newblock \emph{\bibinfo{journal}{Nature}} \textbf{\bibinfo{volume}{578}},
  \bibinfo{pages}{240--245} (\bibinfo{year}{2020}).

\bibitem{QNion}
\bibinfo{author}{Moehring, D.~L.} \emph{et~al.}
\newblock \bibinfo{title}{Entanglement of single-atom quantum bits at a
  distance}.
\newblock \emph{\bibinfo{journal}{Nature}} \textbf{\bibinfo{volume}{449}},
  \bibinfo{pages}{68--71} (\bibinfo{year}{2007}).

\bibitem{QNatom}
\bibinfo{author}{Hofmann, J.} \emph{et~al.}
\newblock \bibinfo{title}{Heralded entanglement between widely separated
  atoms}.
\newblock \emph{\bibinfo{journal}{Science}} \textbf{\bibinfo{volume}{337}},
  \bibinfo{pages}{72--75} (\bibinfo{year}{2012}).

\bibitem{QNNV2}
\bibinfo{author}{Hensen, B.} \emph{et~al.}
\newblock \bibinfo{title}{Loophole-free bell inequality violation using
  electron spins separated by 1.3 kilometres}.
\newblock \emph{\bibinfo{journal}{Nature}} \textbf{\bibinfo{volume}{526}},
  \bibinfo{pages}{682--686} (\bibinfo{year}{2015}).

\bibitem{QNQD}
\bibinfo{author}{Delteil, A.} \emph{et~al.}
\newblock \bibinfo{title}{Generation of heralded entanglement between distant
  hole spins}.
\newblock \emph{\bibinfo{journal}{Nature Physics}}
  \textbf{\bibinfo{volume}{12}}, \bibinfo{pages}{218--223}
  (\bibinfo{year}{2016}).

\bibitem{qcommu}
\bibinfo{author}{Gisin, N.} \& \bibinfo{author}{Thew, R.}
\newblock \bibinfo{title}{Quantum communication}.
\newblock \emph{\bibinfo{journal}{Nature Photonics}}
  \textbf{\bibinfo{volume}{1}}, \bibinfo{pages}{165} (\bibinfo{year}{2007}).

\bibitem{reviewQP}
\bibinfo{author}{Sangouard, N.}, \bibinfo{author}{Simon, C.},
  \bibinfo{author}{de~Riedmatten, H.} \& \bibinfo{author}{Gisin, N.}
\newblock \bibinfo{title}{Quantum repeaters based on atomic ensembles and
  linear optics}.
\newblock \emph{\bibinfo{journal}{Rev. Mod. Phys.}}
  \textbf{\bibinfo{volume}{83}}, \bibinfo{pages}{33--80}
  (\bibinfo{year}{2011}).

\bibitem{satellite}
\bibinfo{author}{Yin, J.} \emph{et~al.}
\newblock \bibinfo{title}{Satellite-based entanglement distribution over 1200
  kilometers}.
\newblock \emph{\bibinfo{journal}{Science}} \textbf{\bibinfo{volume}{356}},
  \bibinfo{pages}{1140--1144} (\bibinfo{year}{2017}).

\bibitem{twophoton}
\bibinfo{author}{Zhao, B.}, \bibinfo{author}{Chen, Z.-B.},
  \bibinfo{author}{Chen, Y.-A.}, \bibinfo{author}{Schmiedmayer, J.} \&
  \bibinfo{author}{Pan, J.-W.}
\newblock \bibinfo{title}{Robust creation of entanglement between remote memory
  qubits}.
\newblock \emph{\bibinfo{journal}{Phys. Rev. Lett.}}
  \textbf{\bibinfo{volume}{98}}, \bibinfo{pages}{240502}
  (\bibinfo{year}{2007}).

\bibitem{nonhiera}
\bibinfo{author}{Sinclair, N.} \emph{et~al.}
\newblock \bibinfo{title}{Spectral multiplexing for scalable quantum photonics
  using an atomic frequency comb quantum memory and feed-forward control}.
\newblock \emph{\bibinfo{journal}{Phys. Rev. Lett.}}
  \textbf{\bibinfo{volume}{113}}, \bibinfo{pages}{053603}
  (\bibinfo{year}{2014}).

\bibitem{multiQP}
\bibinfo{author}{Simon, C.} \emph{et~al.}
\newblock \bibinfo{title}{Quantum repeaters with photon pair sources and
  multimode memories}.
\newblock \emph{\bibinfo{journal}{Phys. Rev. Lett.}}
  \textbf{\bibinfo{volume}{98}}, \bibinfo{pages}{190503}
  (\bibinfo{year}{2007}).

\bibitem{spinwave15}
\bibinfo{author}{G\"undo\ifmmode~\breve{g}\else \u{g}\fi{}an, M.},
  \bibinfo{author}{Ledingham, P.~M.}, \bibinfo{author}{Kutluer, K.},
  \bibinfo{author}{Mazzera, M.} \& \bibinfo{author}{de~Riedmatten, H.}
\newblock \bibinfo{title}{Solid state spin-wave quantum memory for time-bin
  qubits}.
\newblock \emph{\bibinfo{journal}{Phys. Rev. Lett.}}
  \textbf{\bibinfo{volume}{114}}, \bibinfo{pages}{230501}
  (\bibinfo{year}{2015}).

\bibitem{sixhours}
\bibinfo{author}{Zhong, M.} \emph{et~al.}
\newblock \bibinfo{title}{Optically addressable nuclear spins in a solid with a
  six-hour coherence time}.
\newblock \emph{\bibinfo{journal}{Nature}} \textbf{\bibinfo{volume}{517}},
  \bibinfo{pages}{177--180} (\bibinfo{year}{2015}).

\bibitem{band16g}
\bibinfo{author}{Saglamyurek, E.} \emph{et~al.}
\newblock \bibinfo{title}{A multiplexed light-matter interface for fibre-based
  quantum networks}.
\newblock \emph{\bibinfo{journal}{Nature Communications}}
  \textbf{\bibinfo{volume}{7}}, \bibinfo{pages}{11202} (\bibinfo{year}{2016}).

\bibitem{highF}
\bibinfo{author}{Zhou, Z.-Q.}, \bibinfo{author}{Lin, W.-B.},
  \bibinfo{author}{Yang, M.}, \bibinfo{author}{Li, C.-F.} \&
  \bibinfo{author}{Guo, G.-C.}
\newblock \bibinfo{title}{Realization of reliable solid-state quantum memory
  for photonic polarization qubit}.
\newblock \emph{\bibinfo{journal}{Phys. Rev. Lett.}}
  \textbf{\bibinfo{volume}{108}}, \bibinfo{pages}{190505}
  (\bibinfo{year}{2012}).

\bibitem{highF2}
\bibinfo{author}{Zhong, T.} \emph{et~al.}
\newblock \bibinfo{title}{Nanophotonic rare-earth quantum memory with optically
  controlled retrieval}.
\newblock \emph{\bibinfo{journal}{Science}} \textbf{\bibinfo{volume}{357}},
  \bibinfo{pages}{1392--1395} (\bibinfo{year}{2017}).

\bibitem{mode2}
\bibinfo{author}{Tang, J.-S.} \emph{et~al.}
\newblock \bibinfo{title}{Storage of multiple single-photon pulses emitted from
  a quantum dot in a solid-state quantum memory}.
\newblock \emph{\bibinfo{journal}{Nature Communications}}
  \textbf{\bibinfo{volume}{6}}, \bibinfo{pages}{8652} (\bibinfo{year}{2015}).

\bibitem{mode1}
\bibinfo{author}{Zhou, Z.-Q.} \emph{et~al.}
\newblock \bibinfo{title}{Quantum storage of three-dimensional
  orbital-angular-momentum entanglement in a crystal}.
\newblock \emph{\bibinfo{journal}{Phys. Rev. Lett.}}
  \textbf{\bibinfo{volume}{115}}, \bibinfo{pages}{070502}
  (\bibinfo{year}{2015}).

\bibitem{mode3}
\bibinfo{author}{Yang, T.-S.} \emph{et~al.}
\newblock \bibinfo{title}{Multiplexed storage and real-time manipulation based
  on a multiple degree-of-freedom quantum memory}.
\newblock \emph{\bibinfo{journal}{Nature Communications}}
  \textbf{\bibinfo{volume}{9}}, \bibinfo{pages}{3407} (\bibinfo{year}{2018}).

\bibitem{AFC}
\bibinfo{author}{Afzelius, M.}, \bibinfo{author}{Simon, C.},
  \bibinfo{author}{de~Riedmatten, H.} \& \bibinfo{author}{Gisin, N.}
\newblock \bibinfo{title}{Multimode quantum memory based on atomic frequency
  combs}.
\newblock \emph{\bibinfo{journal}{Phys. Rev. A}} \textbf{\bibinfo{volume}{79}},
  \bibinfo{pages}{052329} (\bibinfo{year}{2009}).

\bibitem{qw2}
\bibinfo{author}{G\"uhne, O.} \& \bibinfo{author}{T\'oth, G.}
\newblock \bibinfo{title}{Entanglement detection}.
\newblock \emph{\bibinfo{journal}{Physics Reports}}
  \textbf{\bibinfo{volume}{474}}, \bibinfo{pages}{1 -- 75}
  (\bibinfo{year}{2009}).

\bibitem{QDE4}
\bibinfo{author}{Dousse, A.} \emph{et~al.}
\newblock \bibinfo{title}{Ultrabright source of entangled photon pairs}.
\newblock \emph{\bibinfo{journal}{Nature}} \textbf{\bibinfo{volume}{466}},
  \bibinfo{pages}{217--220} (\bibinfo{year}{2010}).

\bibitem{eleccontrol2}
\bibinfo{author}{Liu, C.} \emph{et~al.}
\newblock \bibinfo{title}{On-demand quantum storage of photonic qubits in an
  on-chip waveguide}.
\newblock \emph{\bibinfo{journal}{Phys. Rev. Lett.}}
  \textbf{\bibinfo{volume}{125}}, \bibinfo{pages}{260504}
  (\bibinfo{year}{2020}).

\bibitem{NdEPR}
\bibinfo{author}{Li, P.-Y.} \emph{et~al.}
\newblock \bibinfo{title}{Hyperfine structure and coherent dynamics of
  rare-earth spins explored with electron-nuclear double resonance at subkelvin
  temperatures}.
\newblock \emph{\bibinfo{journal}{Phys. Rev. Applied}}
  \textbf{\bibinfo{volume}{13}}, \bibinfo{pages}{024080}
  (\bibinfo{year}{2020}).

\bibitem{cavity1}
\bibinfo{author}{Sabooni, M.}, \bibinfo{author}{Li, Q.},
  \bibinfo{author}{Kr\"oll, S.} \& \bibinfo{author}{Rippe, L.}
\newblock \bibinfo{title}{Efficient quantum memory using a weakly absorbing
  sample}.
\newblock \emph{\bibinfo{journal}{Phys. Rev. Lett.}}
  \textbf{\bibinfo{volume}{110}}, \bibinfo{pages}{133604}
  (\bibinfo{year}{2013}).

\bibitem{Er3}
\bibinfo{author}{Ran{\v{c}}i{\'{c}}, M.}, \bibinfo{author}{Hedges, M.~P.},
  \bibinfo{author}{Ahlefeldt, R.~L.} \& \bibinfo{author}{Sellars, M.~J.}
\newblock \bibinfo{title}{Coherence time of over a second in a
  telecom-compatible quantum memory storage material}.
\newblock \emph{\bibinfo{journal}{Nature Physics}}
  \textbf{\bibinfo{volume}{14}}, \bibinfo{pages}{50--54}
  (\bibinfo{year}{2018}).

\setcounter{firstbib}{\value{enumiv}}  
\setcounter{firstbib}{\value{NAT@ctr}}

\end{thebibliography}

\begin{thebibliography}{10}

\setcounter{enumiv}{\value{firstbib}}
\setcounter{NAT@ctr}{\value{firstbib}}

\expandafter\ifx\csname url\endcsname\relax
  \def\url#1{\texttt{#1}}\fi
\expandafter\ifx\csname urlprefix\endcsname\relax\def\urlprefix{URL }\fi
\providecommand{\bibinfo}[2]{#2}
\providecommand{\eprint}[2][]{\url{#2}}

\bibitem{postentanglement}
\bibinfo{author}{Bao, X.-H.} \emph{et~al.}
\newblock \bibinfo{title}{Generation of narrow-band polarization-entangled
  photon pairs for atomic quantum memories}.
\newblock \emph{\bibinfo{journal}{Phys. Rev. Lett.}}
  \textbf{\bibinfo{volume}{101}}, \bibinfo{pages}{190501}
  (\bibinfo{year}{2008}).

\bibitem{tomo}
\bibinfo{author}{James, D. F.~V.}, \bibinfo{author}{Kwiat, P.~G.},
  \bibinfo{author}{Munro, W.~J.} \& \bibinfo{author}{White, A.~G.}
\newblock \bibinfo{title}{Measurement of qubits}.
\newblock \emph{\bibinfo{journal}{Phys. Rev. A}} \textbf{\bibinfo{volume}{64}},
  \bibinfo{pages}{052312} (\bibinfo{year}{2001}).

\bibitem{type2fusion}
\bibinfo{author}{Browne, D.~E.} \& \bibinfo{author}{Rudolph, T.}
\newblock \bibinfo{title}{Resource-efficient linear optical quantum
  computation}.
\newblock \emph{\bibinfo{journal}{Phys. Rev. Lett.}}
  \textbf{\bibinfo{volume}{95}}, \bibinfo{pages}{010501}
  (\bibinfo{year}{2005}).

\bibitem{sdswapping}
\bibinfo{author}{Pan, J.-W.}, \bibinfo{author}{Bouwmeester, D.},
  \bibinfo{author}{Weinfurter, H.} \& \bibinfo{author}{Zeilinger, A.}
\newblock \bibinfo{title}{Experimental entanglement swapping: Entangling
  photons that never interacted}.
\newblock \emph{\bibinfo{journal}{Phys. Rev. Lett.}}
  \textbf{\bibinfo{volume}{80}}, \bibinfo{pages}{3891--3894}
  (\bibinfo{year}{1998}).

\bibitem{twoes2}
\bibinfo{author}{Xu, P.} \emph{et~al.}
\newblock \bibinfo{title}{Two-hierarchy entanglement swapping for a linear
  optical quantum repeater}.
\newblock \emph{\bibinfo{journal}{Phys. Rev. Lett.}}
  \textbf{\bibinfo{volume}{119}}, \bibinfo{pages}{170502}
  (\bibinfo{year}{2017}).

\bibitem{QDE1}
\bibinfo{author}{Huber, D.} \emph{et~al.}
\newblock \bibinfo{title}{Strain-tunable gaas quantum dot: A nearly
  dephasing-free source of entangled photon pairs on demand}.
\newblock \emph{\bibinfo{journal}{Phys. Rev. Lett.}}
  \textbf{\bibinfo{volume}{121}}, \bibinfo{pages}{033902}
  (\bibinfo{year}{2018}).

\bibitem{QDE2}
\bibinfo{author}{Wang, H.} \emph{et~al.}
\newblock \bibinfo{title}{On-demand semiconductor source of entangled photons
  which simultaneously has high fidelity, efficiency, and
  indistinguishability}.
\newblock \emph{\bibinfo{journal}{Phys. Rev. Lett.}}
  \textbf{\bibinfo{volume}{122}}, \bibinfo{pages}{113602}
  (\bibinfo{year}{2019}).

\bibitem{QDE3}
\bibinfo{author}{Liu, J.} \emph{et~al.}
\newblock \bibinfo{title}{A solid-state source of strongly entangled photon
  pairs with high brightness and indistinguishability}.
\newblock \emph{\bibinfo{journal}{Nature Nanotechnology}}
  \textbf{\bibinfo{volume}{14}}, \bibinfo{pages}{586--593}
  (\bibinfo{year}{2019}).

\bibitem{semihiera}
\bibinfo{author}{Liu, X.}, \bibinfo{author}{Zhou, Z.-Q.}, \bibinfo{author}{Hua,
  Y.-L.}, \bibinfo{author}{Li, C.-F.} \& \bibinfo{author}{Guo, G.-C.}
\newblock \bibinfo{title}{Semihierarchical quantum repeaters based on moderate
  lifetime quantum memories}.
\newblock \emph{\bibinfo{journal}{Phys. Rev. A}} \textbf{\bibinfo{volume}{95}},
  \bibinfo{pages}{012319} (\bibinfo{year}{2017}).

\bibitem{jinswapping}
\bibinfo{author}{Jin, J.} \emph{et~al.}
\newblock \bibinfo{title}{Entanglement swapping with quantum-memory-compatible
  photons}.
\newblock \emph{\bibinfo{journal}{Phys. Rev. A}} \textbf{\bibinfo{volume}{92}},
  \bibinfo{pages}{012329} (\bibinfo{year}{2015}).

\bibitem{CHSH}
\bibinfo{author}{Clauser, J.~F.}, \bibinfo{author}{Horne, M.~A.},
  \bibinfo{author}{Shimony, A.} \& \bibinfo{author}{Holt, R.~A.}
\newblock \bibinfo{title}{Proposed experiment to test local hidden-variable
  theories}.
\newblock \emph{\bibinfo{journal}{Phys. Rev. Lett.}}
  \textbf{\bibinfo{volume}{23}}, \bibinfo{pages}{880--884}
  (\bibinfo{year}{1969}).

\bibitem{QME1}
\bibinfo{author}{Clausen, C.} \emph{et~al.}
\newblock \bibinfo{title}{Quantum storage of photonic entanglement in a
  crystal}.
\newblock \emph{\bibinfo{journal}{Nature}} \textbf{\bibinfo{volume}{469}},
  \bibinfo{pages}{508--511} (\bibinfo{year}{2011}).

\bibitem{QME2}
\bibinfo{author}{Saglamyurek, E.} \emph{et~al.}
\newblock \bibinfo{title}{Broadband waveguide quantum memory for entangled
  photons}.
\newblock \emph{\bibinfo{journal}{Nature}} \textbf{\bibinfo{volume}{469}},
  \bibinfo{pages}{512--515} (\bibinfo{year}{2011}).

\bibitem{euspinwave}
\bibinfo{author}{Jobez, P.} \emph{et~al.}
\newblock \bibinfo{title}{Coherent spin control at the quantum level in an
  ensemble-based optical memory}.
\newblock \emph{\bibinfo{journal}{Phys. Rev. Lett.}}
  \textbf{\bibinfo{volume}{114}}, \bibinfo{pages}{230502}
  (\bibinfo{year}{2015}).

\bibitem{Ybnm}
\bibinfo{author}{Ortu, A.} \emph{et~al.}
\newblock \bibinfo{title}{Simultaneous coherence enhancement of optical and
  microwave transitions in solid-state electronic spins}.
\newblock \emph{\bibinfo{journal}{Nature Materials}}
  \textbf{\bibinfo{volume}{17}}, \bibinfo{pages}{671--675}
  (\bibinfo{year}{2018}).

\bibitem{superhyperfine}
\bibinfo{author}{Kindem, J.~M.} \emph{et~al.}
\newblock \bibinfo{title}{Control and single-shot readout of an ion embedded in
  a nanophotonic cavity}.
\newblock \emph{\bibinfo{journal}{Nature}} \textbf{\bibinfo{volume}{580}},
  \bibinfo{pages}{201--204} (\bibinfo{year}{2020}).

\bibitem{Erkramers}
\bibinfo{author}{Rakonjac, J.~V.}, \bibinfo{author}{Chen, Y.-H.},
  \bibinfo{author}{Horvath, S.~P.} \& \bibinfo{author}{Longdell, J.~J.}
\newblock \bibinfo{title}{Long spin coherence times in the ground state and in
  an optically excited state of
  $^{167}\mathrm{Er}^{3+}:\mathrm{Y}_{2}\mathrm{SiO}_{5}$ at zero magnetic
  field}.
\newblock \emph{\bibinfo{journal}{Phys. Rev. B}}
  \textbf{\bibinfo{volume}{101}}, \bibinfo{pages}{184430}
  (\bibinfo{year}{2020}).

\bibitem{YbPRL}
\bibinfo{author}{Businger, M.} \emph{et~al.}
\newblock \bibinfo{title}{Optical spin-wave storage in a solid-state hybridized
  electron-nuclear spin ensemble}.
\newblock \emph{\bibinfo{journal}{Phys. Rev. Lett.}}
  \textbf{\bibinfo{volume}{124}}, \bibinfo{pages}{053606}
  (\bibinfo{year}{2020}).

\bibitem{Ndeff}
\bibinfo{author}{Bussi{\`e}res, F.} \emph{et~al.}
\newblock \bibinfo{title}{Quantum teleportation from a telecom-wavelength
  photon to a solid-state quantum memory}.
\newblock \emph{\bibinfo{journal}{Nature Photonics}}
  \textbf{\bibinfo{volume}{8}}, \bibinfo{pages}{775--778}
  (\bibinfo{year}{2014}).

\bibitem{widebandeff2}
\bibinfo{author}{Puigibert, M. l.~G.} \emph{et~al.}
\newblock \bibinfo{title}{Entanglement and nonlocality between disparate
  solid-state quantum memories mediated by photons}.
\newblock \emph{\bibinfo{journal}{Phys. Rev. Research}}
  \textbf{\bibinfo{volume}{2}}, \bibinfo{pages}{013039} (\bibinfo{year}{2020}).

\bibitem{TBP}
\bibinfo{author}{Lvovsky, A.~I.}, \bibinfo{author}{Sanders, B.~C.} \&
  \bibinfo{author}{Tittel, W.}
\newblock \bibinfo{title}{Optical quantum memory}.
\newblock \emph{\bibinfo{journal}{Nature Photonics}}
  \textbf{\bibinfo{volume}{3}}, \bibinfo{pages}{706--714}
  (\bibinfo{year}{2009}).

\bibitem{higheff}
\bibinfo{author}{Hedges, M.~P.}, \bibinfo{author}{Longdell, J.~J.},
  \bibinfo{author}{Li, Y.} \& \bibinfo{author}{Sellars, M.~J.}
\newblock \bibinfo{title}{Efficient quantum memory for light}.
\newblock \emph{\bibinfo{journal}{Nature}} \textbf{\bibinfo{volume}{465}},
  \bibinfo{pages}{1052--1056} (\bibinfo{year}{2010}).

\end{thebibliography}
\end{document}